\documentclass[12pt,preprint]{aastex}

\newcommand{\msun}{M$_{\sun}$}

\slugcomment{}

\shorttitle{HI in NGC 5746}
\shortauthors{Rand, Benjamin}

\begin{document}


\title{Vertically Extended Neutral Gas in the Massive Edge-on Spiral NGC 5746}


\author{Richard J. Rand}
\affil{Department of Physics and Astronomy, University of New
Mexico, 800 Yale Blvd, NE, Albuquerque, NM 87131}
\email{rjr@phys.unm.edu}
\and

\author{Robert. A. Benjamin}
\affil{Department of Physics, University of Wisconsin at Whitewater,
800 West Main Street, Whitewater, WI 53190}
\email{ benjamir@uww.edu}


\begin{abstract}

We present Very Large Array 21-cm observations of the massive edge-on spiral
galaxy NGC 5746.  This galaxy has recently been reported to have a luminous
X-ray halo, which has been taken as evidence of residual hot gas as predicted
in galaxy formation scenarios.  Such models also predict that some of this gas
should undergo thermal instabilities, leading to a population of warm clouds
falling onto the disk.  If so, then one might expect to find a vertically
extended neutral layer.  We detect a substantial high-latitude component, but
conclude that almost all of its mass of $1.2-1.6 \times 10^9$ \msun\ most
likely resides in a warp.  Four features far from the plane containing about
$10^8$ \msun\ are found at velocities distinct from this warp.  These clouds
may be associated with the expected infall, although an origin in a disk-halo
flow cannot be ruled out, except for one feature which is counter-rotating.
The warp itself may be a result of infall according to recent models.  But
clearly this galaxy lacks a massive, lagging neutral halo as found in NGC 891.
The disk HI is concentrated into two rings of radii 1.5 and 3 arcminutes.
Radial inflow is found in the disk, probably due to the bar in this
galaxy.  A nearby member of this galaxy group, NGC 5740, is also detected.  It
shows a prominent one-sided extension which may be the result of ram pressure
stripping.

\end{abstract}





\section{Introduction}

Gaseous thick disks or halos in spiral galaxies hold promise for answering key
questions about galaxy formation and evolution.  These vertically extended
layers are multiphase, with detailed studies of the nearest edge-on galaxies
revealing halos of neutral hydrogen \citep[e.g.][]{1994ApJ...429..618I,
1997ApJ...491..140S}, diffuse ionized gas \citep[DIG; e.g.][]{
1990ApJ...352L...1R, 1990A&A...232L..15D, 2003A&A...406..505R}, hot X-ray
emitting gas \citep[e.g.][]{1994ApJ...420..570B, 2004ApJ...606..829S,
2006A&A...448...43T}, radio continuum emission \citep[e.g.][]{
1999AJ....117.2102I,2001A&A...374...42D} and dust
\citep[e.g.][]{1999AJ....117.2077H,2000A&AS..145...83A, 2006A&A...445..123I}.

Especially for the extended DIG, X-ray, dust and radio continuum components,
it has been well established that their prominence is correlated with the star
formation activity in the underlying disk \citep[e.g.][]{1996ApJ...462..712R,
2003A&A...406..493R,1999AJ....117.2077H,2006A&A...457..779T,2006A&A...457..121D}.
Large shell-like and vertically oriented filamentary structures are also seen
in many of the more actively star-forming edge-ons, in, e.g., DIG \citep
{1990ApJ...352L...1R} and HI \citep[e.g.][]{2001A&A...377..759L}.  All of this
evidence has led to a picture of the origin of these layers in a
star-formation-driven disk-halo cycle, which has been theoretically modeled as
a general galactic fountain flow
\citep{1976ApJ...205..762S,1980ApJ...236..577B} with later models
incorporating the fact that mass and energy input into the halo may
efficiently occur through localized structures such as supershells and
chimneys \citep{1989ApJ...345..372N}.

As for the Milky Way, the WHAM survey \citep{2003ApJS..149..405H} has
characterized in detail the so-called Reynolds layer of vertically-extended
DIG \citep{1973ApJ...179..651R}, revealing further instances of possible
superbubbles \citep{2001ApJ...558L.101R}.  Very relevant to this paper are the
High Velocity Clouds (HVCs) and Intermediate Velocity Clouds (IVCs).  For the
latter, a few are known to be 0.3--4 kpc above the midplane, with
metallicities close to solar, and it is quite possible that these clouds
originate in a disk-halo flow (\citealt{2004Ap&SS.289..381W}, and references
therein).  The situation for the HVCs is different, however.  Not
considering the contribution from the Magellanic Stream
\citep{2003ApJ...586..170P}, some well studied complexes are found to be many
kpc from the plane (e.g. 8--10 kpc for part of complex A, $>$4 kpc for complex
C; \citealt {2004Ap&SS.289..381W} and references therein).  Although
information is scarce, their metallicities are also well below solar
(e.g. \citealt{2007ApJ...657..271C}), suggesting that such clouds are not part
of a disk-halo flow but may be extragalactic clouds that are mixing with
metal-enriched gas in the halo as they fall onto the disk.

Such a population of infalling warm clouds is a prediction of recent models of
galaxy formation, in which galaxies are still growing by this mechanism.  The
idea that galaxies form by gas cooling out of a shock-heated hot halo was
presented by \citet{1978MNRAS.183..341W}.  This hot gas should be thermally
unstable and given to fragmentation \citep{1965ApJ...142..531F,
1985ApJ...298...18F,1990ApJ...363...50M}.  \citet{2004MNRAS.355..694M}
explored the consequences of such an inflow of fragmenting material.  The
remaining hot gas has such a low density that it can support itself against
infall for a long time.  This may explain the "over-cooling" problem in galaxy
formation models that leads to overly massive galaxies and the finding that
most of the baryons in the universe never ended up in galaxies
(\citealt{2003ApJ...599...38B} and references therein).  Instabilities in
cooling inflows have also been studied by \citet{2006MNRAS.370.1612K} and
\citet{2006ApJ...644L...1S}.  The cloud populations and distributions in these
high-resolution models vary, with clouds having an uncertain but probably high
ionization fraction.

An alternative explanation to the "over-cooling" problem in the most massive
galaxies is feedback.  Reheating of the halo gas by core-collapse supernovae
or a (possibly recurrent) AGN prevents the gas from cooling and flowing in
\citep{2004MNRAS.347.1093B}.  Heating by Type Ia supernovae is considered by
\citet{2005ASPC..331..329W}.  It is also argued that, except in the most
massive halos, infalling gas never heats to the virial temperature but flows
in cold \citep{1977ApJ...215..483B,2003MNRAS.345..349B,2004MNRAS.347.1093B,
2005MNRAS.363....2K,2006MNRAS.368....2D}.


The rotation of gaseous halos may also provide clues as to their origin and
the physical processes occurring within them.  In recent years, the manner in
which rotation speeds change as a function of height has begun to be
characterized in the DIG \citep{2006ApJ...647.1018H, 2006ApJ...636..181H,
2007ApJ...663..933H, 2007AJ....134.1019O}.  Heald and coworkers have measured
the gradient in rotation speed with height ($dV_{rot}/dz$) in three edge-ons
which form a decreasing sequence of star forming activity and DIG
scale-height: NGC 5775 (--8 km s$^{-1}$ kpc$^{-1}$), NGC 891 (--17 km s$^{-1}$
kpc$^{-1}$), and NGC 4302 (--30 km s$^{-1}$ kpc$^{-1}$).  The gradients, when
expressed in terms of km s$^{-1}$ per DIG scale-height, show much less range:
--15 to --25 km s$^{-1}$ (scale height)$^{-1}$.  The authors show that simple
ballistic models of disk-halo flow \citep{2002ApJ...578...98C} greatly
underpredict the magnitude of these gradients and also predict the wrong trend
with scale-height.  Two possible resolutions are: 1) the physics of disk-halo
flows are not well described by ballistic models and that hydrodynamical
effects such as pressure \citep{2006A&A...446...61B} or magnetic or viscous
forces \citep{2002ASPC..276..201B} dominate the rotation, or 2) part or all of
the gas does not originate in a flow from the high-angular-momentum disk, but
rather from infalling low-angular-momentum primordial gas.  For the DIG, the
latter explanation must also account for morphological and other connections
with ongoing star formation discussed above.

For NGC 891, $dV_{rot}/dz$ has also been measured for the HI and is --15 km
s$^{-1}$ kpc$^{-1}$ \citep{2007AJ....134.1019O}, in good agreement with the
DIG value .  The authors conclude from this gradient that at least some of the
$1.2 \times 10^9$ \msun\ HI halo must be accreted, low angular momentum
material (see also \citealt{2006MNRAS.366..449F}).

Furthering our understanding of both galactic infall and disk-halo cycling, as
well as possible interactions between the two, would be made easier if one
could isolate galaxies where one or the other origin is expected to obtain.
High star-forming galaxies like NGC 5775 presumably have gaseous halos
dominated by disk-halo flows.  The recent discovery of a bright ( $L_x = 7.3
\pm 3.9 \times 10^{39}$ erg s$^{-1}$) X-ray halo around the massive, nearby
(29.4 Mpc is the commonly adopted distance), low star forming edge-on Sb
galaxy NGC 5746 \citep{2006NewA...11..465P,2006astro.ph.10893R} presents a
challenge to disk-halo models as the X-ray luminosity clearly exceeds that
expected \citep{2006A&A...457..779T,2004ApJ...606..829S} for a galaxy with no
detected DIG halo (\citealt{1996ApJ...462..712R}; in an image with an Emission
Measure noise level of 3.7 pc cm$^{-6}$) and little star formation.
\citet{1996ApJ...462..712R} roughly characterized the star formation rate per
unit disk area of many edge-ons using the far infrared luminosity measured by
the {\it Infrared Astronomical Satellite} (IRAS) divided by the optical disk
area: $L_{FIR}$/$D_{25}^2$ (values calculated from the NASA/IPAC Extragalactic
Database\footnote{The NASA/IPAC Extragalactic Database (NED) is operated by
the Jet Propulsion Laboratory, California Institute of Technology, under
contract with the National Aeronautics and Space Administration.}).  NGC 5746
is one of the lower of the 16 edge-ons thus characterized, at $4 \times
10^{39}$ erg s$^{-1}$ kpc$^{-2}$.  Rather, the X-ray luminosity puts NGC 5746
on the expected steep relationship $L_x \propto V_{rot}^7$ for hot halos in
galaxy formation models \citep{2002MNRAS.335..799T}, and, in such an
interpretation, it is only because of the very high rotation speed (measured
in this paper to be about 310 km s$^{-1}$) that this residual hot halo can be
detected at all.  The metallicity of the hot gas is found to be low, at about
0.04 solar \citep{2006astro.ph.10893R}, but is uncertain and could be biased
toward low values (see \citealt{2000MNRAS.311..176B}).  But overall, this
galaxy may well be an attractive test case where disk-halo cycling is
minimized.

There are some caveats to this interpretation, however.  First, NGC 5746 is in
a group of 26 cataloged members \citep{2000ApJ...543..178G}, and there may
well have been interactions in the past leading to gas in the group
environment.  Second, although there is little ongoing star formation, it may
be that Type Ia supernovae or an AGN create a hot
wind or provide energy to keep a pre-existing halo heated, as discussed above.
However, \citet{2006astro.ph.10893R} argue that the likely Type Ia SN rate in
a galaxy like NGC 5746 would be far too low to explain the X-ray luminosity.
They also argue that there is little evidence for at least current nuclear
activity in this galaxy, and that a soft, thermal X-ray halo would not be
expected for an AGN outflow anyway.  We do note, however, that NGC 5746 is
classified as a LINER by \citet{1999RMxAA..35..187C}, and there is a compact
X-ray source \citep{2006A&A...460...45G} at the center, indicating at least
low level activity at present, although no compact radio source has been
detected \citep{2005A&A...435..521N}.  If the metallicity of the hot gas is
indeed as low as 0.04, these alternative sources are not very likely.

If this X-ray halo is a residual of galaxy formation and thus an indication of
missing baryons, then an obvious question is whether the predicted thermal
instabilities are occurring, leading to an infalling warm component.  If it
exists, the lack of a DIG halo may simply mean that there are few disk sources
capable of significant ionization.  But there may be a neutral halo.  Either a
detection or an upper limit will constrain galaxy formation models and
the \lq\lq missing baryon\rq\rq\ question, and shed light on the origin of the
Milky Way's HVCs, more so if individual clouds can be seen.  A detection of a
significant neutral halo would also be challenging to explain in terms of a
disk-halo cycle origin.

We have therefore observed NGC 5746 in 21-cm emission with the Very Large
Array (VLA), as described in \S 2.  We analyze high-latitude emission in
\S 3, and end with a brief discussion in \S 4 in terms of the theories of the
halo gas discussed above.

\section{Observations}

NGC 5746 was observed in the C array of the VLA on 2007 January 7, 8, 12 and
13.  Phase calibration was achieved through observations of VLA calibrator
1445+099 about every 30 minutes.  Observations of 3C286 and 3C48 were used for
flux and bandpass calibration.  Sixty three spectral channels were employed,
centered at 1725 km s$^{-1}$, with channel width 20.85 km s$^{-1}$, while
online Hanning smoothing yielding a velocity resolution of twice this width.
A total of about 28.3 hours were spent observing NGC 5746.  Eight of the 27
antennae were unavailable for most of the final track, while two were
absent on the other three dates.  Smaller amounts of data were also lost due
to high winds, interference, and equipment failures.  Data were inspected for
high amplitudes and any suspect data excised.  Some small baseline corrections
for a few antennae were made, based on later observations in the same
configuration.  The calibration of all four tracks is of very high quality.

High amplitude visibilities were clipped before mapping.  Dirty maps of each
track were made to check for additional problems before concatenating the four
$uv$ datasets.  Continuum was subtracted in the $uv$-plane using line-free
channels at either end of the spectrometer with the AIPS tasks UVSUB and
UVLSF.  This was very successful.  The AIPS task IMAGR was used to produce
clean maps using clean boxes that covered the emission from NGC 5746 as well
as NGC 5740, which is also detected, albeit somewhat outside the 31.9' FWHM
primary beam.  The pointing center is R.A. 14$^{\rm h}$ 44$^{\rm m}$
56.4$^{\rm s}$, Decl. 1$^\circ$ 57' 16" (J2000).  The primary data cube
discussed here was made with no $uv$-tapering and has a resolution of
15.5"x14.8" (2.2 x 2.1 kpc) at a P.A. of $-3.6^\circ$.  A cube with
61.3"x58.1" resolution (8.7 x 8.3 kpc), at a P.A. of 45.5$^\circ$, created via
$uv$-tapering is also discussed here.  All maps have 1024x1024 pixels of 3"
size, and were made with uniform $uv$-weighting with the IMAGR 'ROBUST'
parameter equal to 0.  The noise in a single channel of the full-resolution
cube is 0.23 mJy (beam)$^{-1}$.  No primary beam correction was made for maps
of NGC 5746.  Primary beam attenuation reaches 4\% at the ends of the major
axis.  Maps of NGC 5740 are corrected for primary beam attenuation; the
response at the center of this galaxy has dropped to 0.36, and thus structure
may not be accurately mapped.  No other galaxies in the NGC 5746 group are
within the mapped region.  Observed velocities are on the heliocentric scale.
The noise corresponds to an HI column density of $4.6 \times 10^{19}$
cm$^{-2}$ averaged over one resolution element for optically thin gas.  In a
zeroth-moment map made using all data and all channels, a 5$\sigma$ detection
of a point source corresponds to $2.3 \times 10^6$ \msun, using the conversion
from (e.g.)  \citet{1997ApJ...490..173Z}.  All masses are of total atomic gas
and include a correction of 1.36 for helium content.

Finally, the GIPSY task BLOT was used to blank emission-free regions in each
channel, reducing the noise in moment maps.  Moment maps were made with
various strategies to eliminate as much noise as possible while retaining as
much emission as possible that appears to be real in the channel maps and
position-velocity (pv) diagrams (given its signal-to-noise ratio and continuity
over multiple velocity channels).  The zeroth-moment map presented here for
the full-resolution cube includes all emission in the blotted cube above the
1$\sigma$ level in each pixel in two consecutive channels.  For the 60"
resolution cube, such a strategy introduced too many features in the
zeroth-moment map that did not appear real in the data cube, and thus a
2.5$\sigma$ cutoff was used.  A concern with the velocity resolution of 42 km
s$^{-1}$ is that many real spectrally narrow features may be rejected by
this strategy.

\section{Results}

Channel maps of the full-resolution cube are shown in Figure 1.  Zeroth- and
first-moment maps made from the blotted full-resolution cube are shown in
Figure 2.  The zeroth-moment map is overlaid on an Digitized Sky Survey red
image in Figure 3.  Also labeled in Figure 3 are the four extraplanar features
discussed in \S 3.5.  The emission from NGC 5746 is dominated by what appears
to be a highly inclined disk at a position angle (PA) of 350$^\circ$ (measured
CCW from N to the receding side of the major axis) with a large central
depression of about 2' diameter.  In this sense it is somewhat reminiscent of
the {\it Infrared Space Observatory} ISOCAM 12$\mu$m image from
\citet{2002AJ....123.3067B}, which has the appearance of an inclined ring of
approximate diameter 2.3' (19.7 kpc).  It is possible the HI hole is larger
but limited resolution has made it appear smaller.

\clearpage
\begin{figure}
\epsscale{1.00}
\includegraphics[scale=1.0,viewport=17 121 578 697]{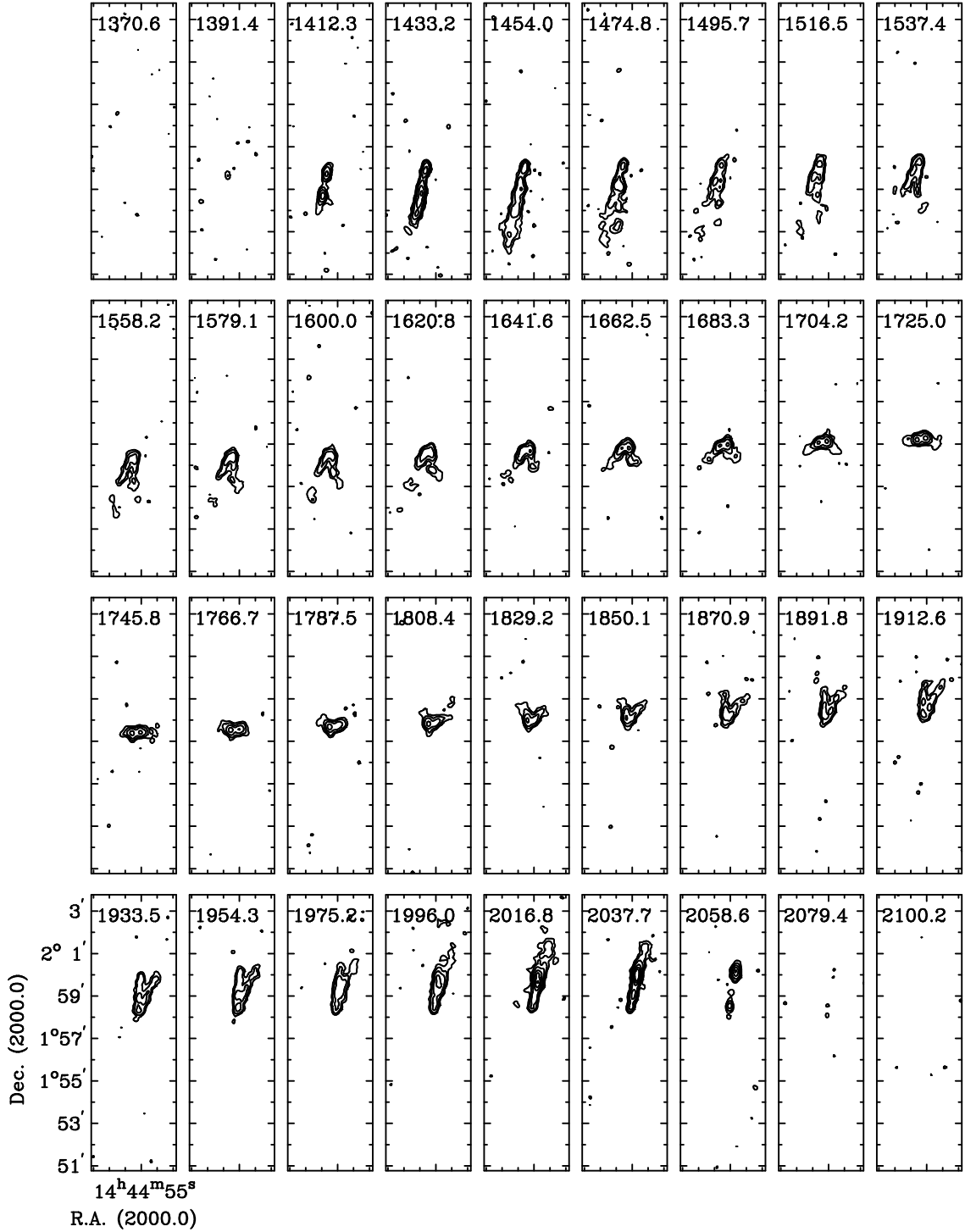}
\caption{Channel maps from the full-resolution cube for NGC 5746.  Contour
levels are 3, 6, 12, 24, 36 and 48 times the 1 $\sigma$ noise in column
density units of $4.6 \times 10^{19}$ cm$^{-2}$.  Each map is labeled with
its heliocentric velocity.
\label{fig1}}
\end{figure}

\begin{figure} \epsscale{1.00}
\includegraphics[scale=1.0,viewport=38 159 480 763]{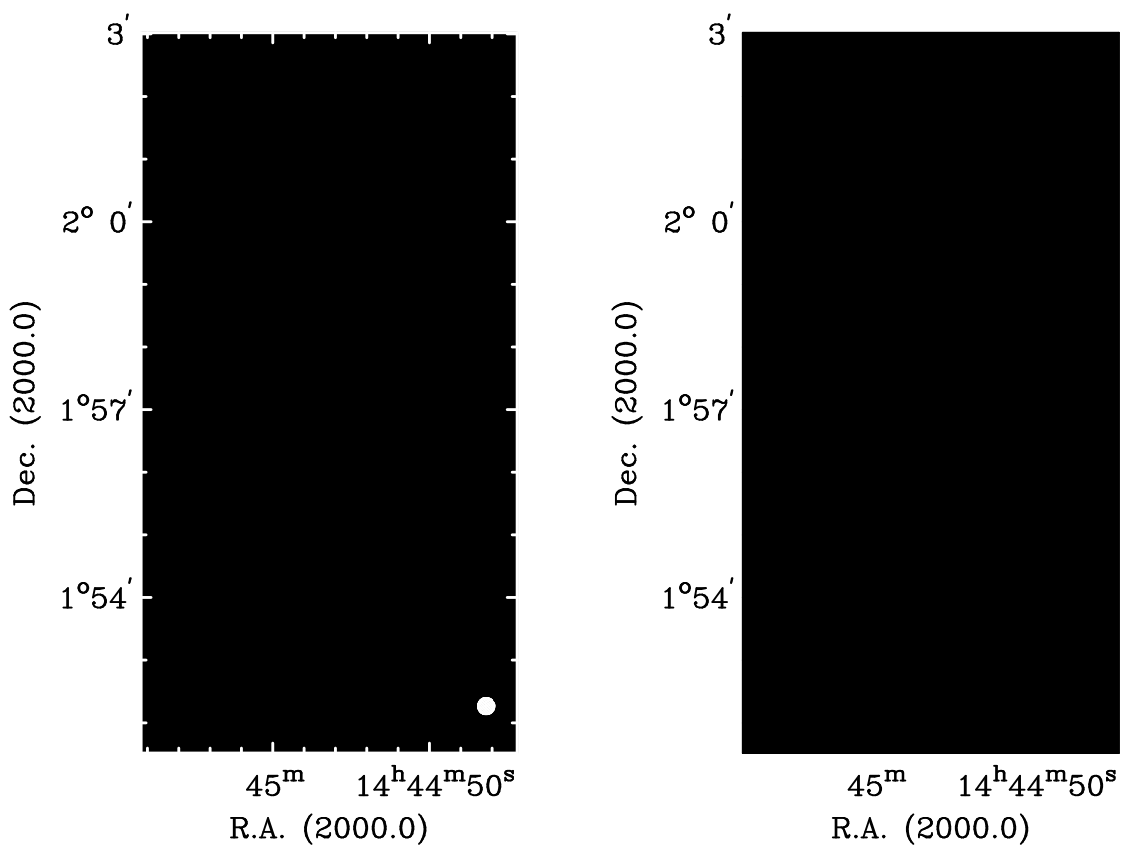}
\caption{{\it Left:} Zeroth-moment map of NGC 5746 from the full-resolution
cube.  Contour levels are 1, 2, 4, 8, 17 and 32 times $10^{20}$ cm$^{-2}$.
The beam is shown in the lower right corner.
{\it Right:} First-moment map from the full-resolution cube.  Contours run from
1500 to 2020 km s$^{-1}$ in 40 km s$^{-1}$ increments from south to north.
\label{fig2}}
\end{figure}

\begin{figure}
\epsscale{0.60}
\includegraphics[scale=1.0,viewport=17 21 578 637]{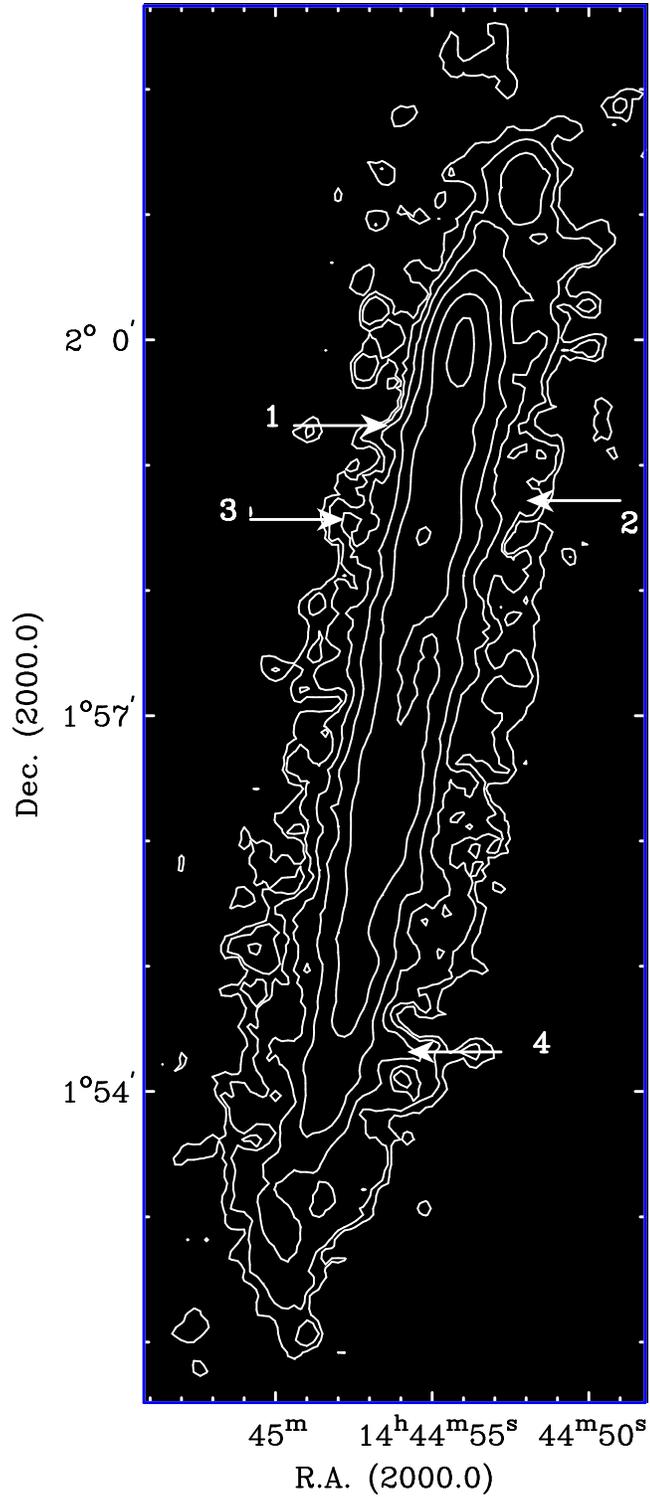}
\caption{Zeroth-moment map of NGC 5746 from the full-resolution
cube overlaid on a Digitized Sky Survey red image.  Contour
levels are as in Figure 2 (left).  The approximate locations of the centers
of the four features discussed in \S 3.5 are also indicated.
\label{fig3}}
\end{figure}
\clearpage

At fainter levels, a component extended along the minor axis is clearly
detected.  It is resolved into individual clouds and filaments to some degree.
This is in contrast to the HI map of NGC 891 (see Figure 1 of
\citealt{2007AJ....134.1019O}) where the halo is smoother in appearance
despite better linear resolution (1.4 kpc vs. 2.1 kpc).  Emission is detected
up to about 70" (10 kpc) from the midplane.  The northern end of the disk
also suggests a warp. 

The total flux in the map in Figure 2 is 33.7 Jy km s$^{-1}$, yielding a total
atomic mass of $9.4 \times 10^9$ \msun.  For comparison, the total observed
flux found by \citet{2005ApJS..160..149S} using the Arecibo telescope is 38.8
Jy km s$^{-1}$.

Figure 4 shows the zeroth- and first-moment maps from the 60"-resolution cube.
The total mass in this map is $8.8 \times 10^9$ \msun, slightly less than in
the full-resolution map.  We attribute this to the somewhat higher cutoff that
was necessary to eliminate noisy features in the moment map at this
resolution.  However, it also indicates that there is very little diffuse, low
surface brightness emission that is missed in the full resolution cube.  The
ends of the disk in this map also suggest a warp, but with the opposite sense
of bending than is evident at lower radii in the full resolution map.

\clearpage
\begin{figure} \epsscale{1.0}
\includegraphics[scale=1.0,viewport=38 159 480 763]{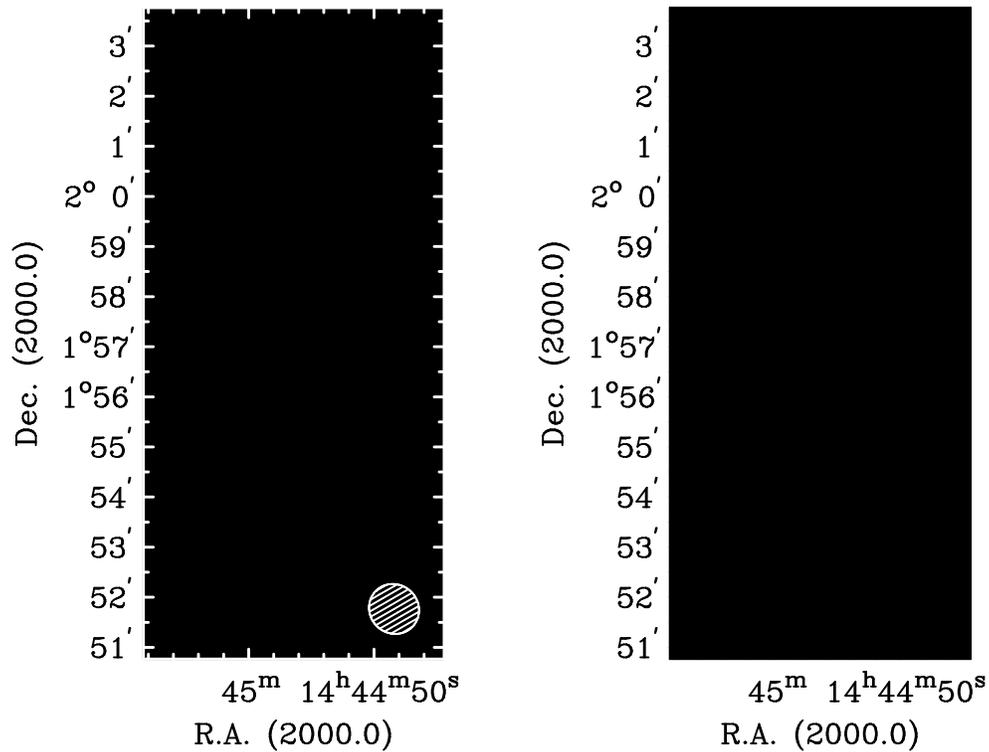} 
\caption{{\it Left:} Zeroth-moment map of NGC 5746 from the 1'-resolution cube.
Contour levels are 1, 2, 4, 8, 16, 32, 64 and 92 times $1.3 \times 10^{19}$ 
cm$^{-2}$.  The beam is shown in the lower right corner.
{\it Right:} First-moment map from the 1'-resolution cube.  Contours run from
1460 to 2020 km s$^{-1}$ in 40 km s$^{-1}$ increments from south to north.
\label{fig4}} \end{figure}
\clearpage

\subsection{Modeling the Atomic Mass Distribution and Rotation}

Our main goal here is to understand the origin of the faint, vertically
extended component seen in Figure 2 through full modeling of the data cubes.
Although we are able to quantitatively constrain many of the model parameters,
we do not carry out a full search of parameter space, not least because of
asymmetries in the data, but rather explore and constrain a few well motivated
types of models.

We begin by constraining the radial density profile at the midplane.  As a
first estimate, we employ the GIPSY task RADIAL, which fits radial profiles of
the column density integrated vertically through the disk to averaged major
axis emission profiles for highly inclined galaxies.  The observed major axis
emission profile is shown in Figure 5 as the solid line.  Since in the below
modeling we will use axisymmetric density distributions, we also show a
profile where the north and south sides have been averaged to produce an
axisymmetric profile (dashed line).  The fitted axisymmetric profile is shown
as the dotted line.  The radial column density profile derived by RADIAL is
shown in Figure 6 as the filled circles.  The disk is dominated by two rings
of radius 1.5' and 3' and width about 1', while the profile confirms the
aforementioned central depression of radius 1'.  RADIAL reproduces the
observed profile in Figure 5 very well, except for a moderate overestimate of
the emission in the central depression.  This overestimate cannot be rectified
by creating a more prominent central hole but is instead a result of projected
emission from gas at larger radii; it presumably indicates an asymmetry in the
galaxy.  The other symbols in Figure 6 show the deviations from the basic
profile featured in the models described below.

\clearpage
\begin{figure} \epsscale{1.0}
\includegraphics[scale=1.0,viewport=107 41 578 597]{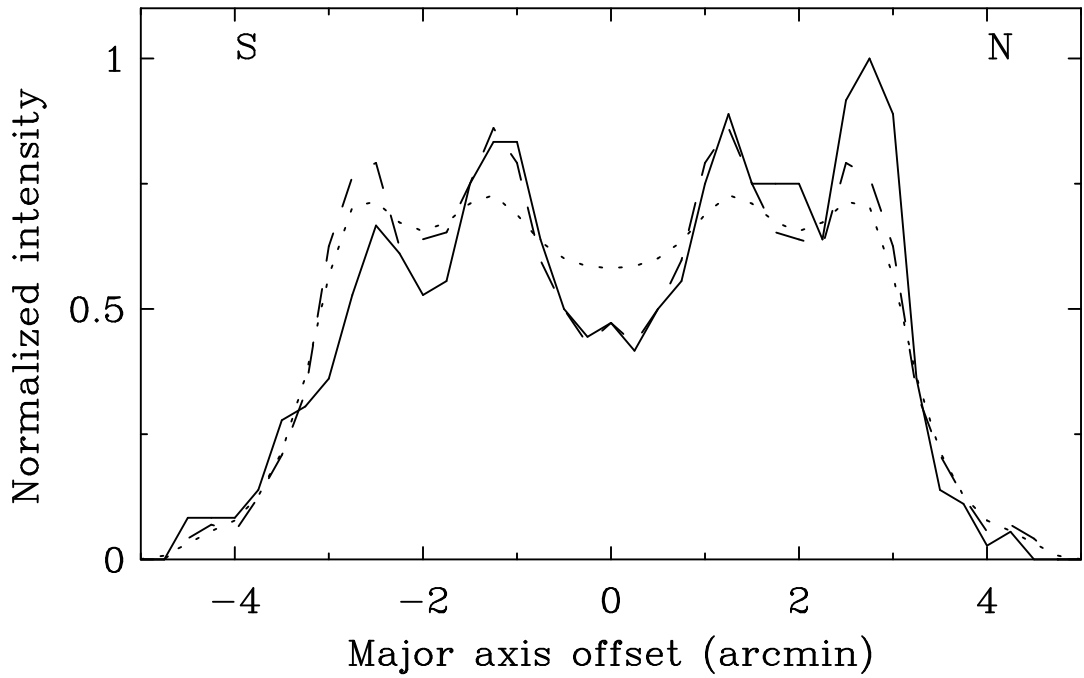} 
\caption{The solid line shows the observed major axis emission profile
averaged over the minor-axis extent of the emission in NGC 5746.  The dashed
line shows the average of the north and south sides of the observed profile.
The dotted line shows the fitted profile from the GIPSY task RADIAL.
\label{fig5}} \end{figure}

\begin{figure} \epsscale{1.0}
\includegraphics[scale=1.0,viewport=7 41 578 597]{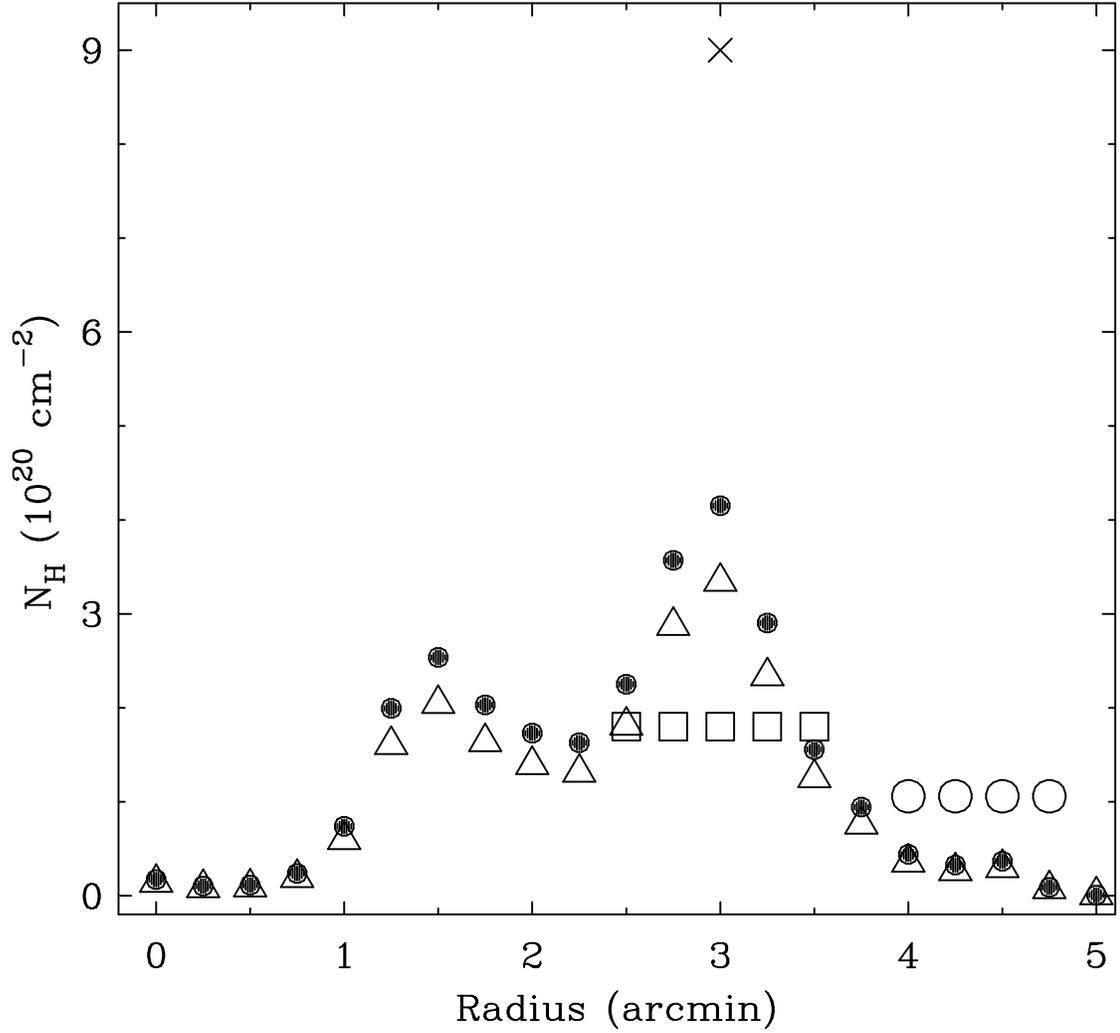}
\caption{Filled circles show the axisymmetric radial profile
of the column densities integrated vertically through the disk derived by
RADIAL for NGC 5746.  This profile is used for the unwarped component of
Models A and B.  Open circles show the modification necessary to model the
extended emission along the minor axis as a warp in Models A and B.  Triangles
show the slightly lower profile for the disk component in the lagging halo
Models C, D, and E.  The 'x' and the open squares show the column density of
the ringlike lagging halo Models D and E, respectively.  See text for further
details of the models.
\label{fig6}} \end{figure}
\clearpage

We use the major axis pv diagram (Figure 7) and the envelope tracing method to
estimate the rotation curve.  The method is described in
\citet{2001ARA&A..39..137S}.  We use $\eta=0.3$ (see
\citealt{2001ARA&A..39..137S} for the definition of this parameter).  We
assume the inclination is close enough to $90^\circ$ so that no correction for
projection is necessary.  As there is little modeled column density in the
central 1' radius, the rotation curve is very poorly constrained there.  In
fact, long-slit spectra and a peanut bulge indicate that NGC 5746 is clearly
barred, with the bar elongated more across the line of sight than along it
\citep{1995ApJ...443L..13K, 1999AJ....118..126B}.  The bar manifests itself
kinematically in a steeply rising component within major axis offsets of $\pm
0.25'$ in pv diagrams of emission lines (indicative of $x_2$ orbits
perpendicular to the bar), which is clearly not seen in HI.  For our rotation
curve we have simply extended the value at $R=1.15'$ inwards.  Our adopted
rotation curve is shown in Figure 8.  The dynamical mass within $R=5'$ (42.8
kpc), given by $M(R) = R\,V^2(R)/G$, is $7.9 \times 10^{11}$ \msun.

\clearpage
\begin{figure} \epsscale{1.0}
\includegraphics[scale=1.0,viewport=27 241 578 797]{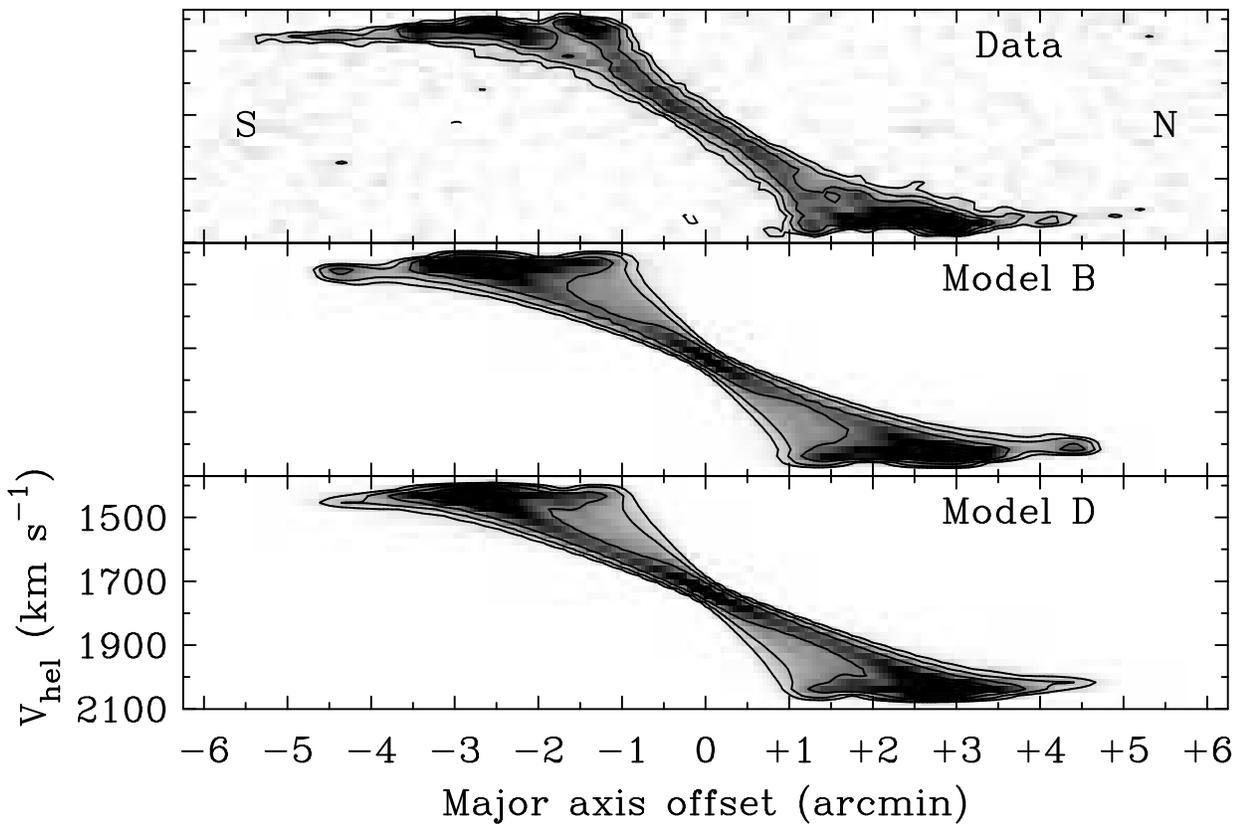} 
\caption{At the top, middle and bottom are shown major-axis position-velocity
diagrams for NGC 5746 from the full-resolution cube, Model B and Model D
respectively (see text for model details).
Contour levels are as in Figure 1.
\label{fig7}} \end{figure}

\begin{figure} \epsscale{1.0}
\includegraphics[scale=1.0,viewport=77 41 578 547]{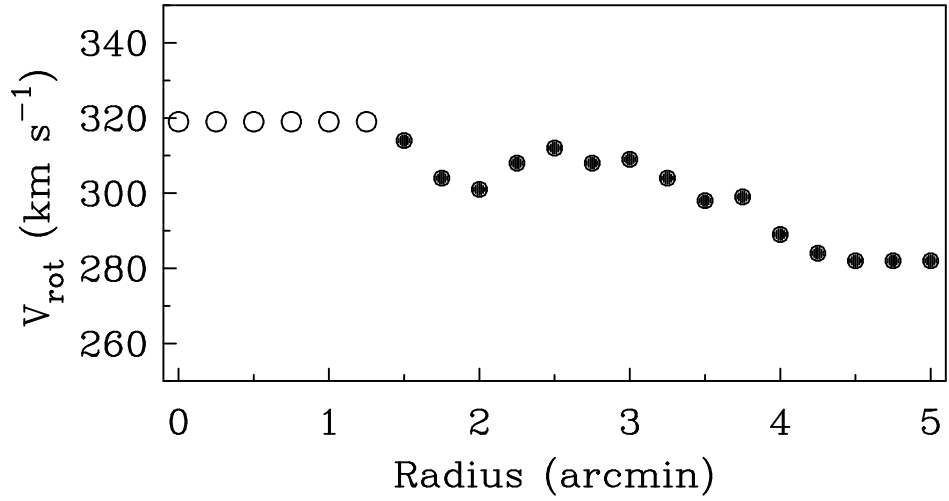} 
\caption{The filled circles show the rotation curve for NGC 5746 derived from
envelope fitting.  The open circles show the assumed extension at
low radii.
\label{fig8}} \end{figure}
\clearpage

We next need to constrain the inclination, $i$.  For nearly edge-on galaxies,
tilted-ring programs such as ROTCUR in GIPSY cannot do this accurately,
especially if a warp along the line of sight or a lagging halo is suspected
and contributes significantly to the velocity field.  However, $i$ can be
estimated by assuming axisymmetry in the ringlike appearance of the emission
at the center of the zeroth-moment map.  Hence, we use our adopted rotation
curve and radial column density profile as inputs to the GIPSY program GALMOD
(a program which allows construction of model galaxies with specified radial
column density profiles, rotation curves, and various forms for the vertical
density distribution) in order to generate a zeroth-moment map and match it to
the data.  The radial bins in all our models have a width of 15" (2.1 kpc).
The models produced by GALMOD are convolved to the resolution of the data.
From the measured separation of the two sides of the ring along the minor axis
near the galaxy center in Figure 2, we find that $i$ must be about 86$^\circ$.
This value is confirmed by the modeling described below.  ROTCUR was run to
constrain other parameters, and indicates a PA of $350 \pm 0.3^\circ$, a
systemic velocity of $1733 \pm 7$ km s$^{-1}$, and a kinematic center about 6"
or 850 pc west of the pointing center, at R.A. 14$^{\rm h}$ 44$^{\rm m}$
56.0$^{\rm s}$, Decl. 1$^\circ$ 57' 16" (J2000).  The gas is initially assumed
to be in a single exponential layer with scale height 3" or about 400 pc.  The
true scale height of the bulk of the emission is not well constrained because
of the resolution and could be substantially lower than 3''.

We next show that the vertical structure cannot be fit by a single component
at a given inclination, or by two exponential components.  Figure 9 shows a
minor-axis emission profile, averaged over 3' along the major axis (solid
line).  There are clearly tails at both ends of the profile.  The dashed line
shows our modeled component for an exponential layer with 3" scale height.  It
fits the bulk of the emission well but not the tails.  The dotted line shows
an exponential model with a scale height of 18".  It fits the tails well but
is broader than the bulk of the emission and washes out the splitting at low
latitudes (which is due to the ringlike distribution).  No combination of two
exponentials were found to produce a reasonable fit.  This result does not
change if a gaussian or sech$^2$ form is used.

\clearpage
\begin{figure} \epsscale{.80}
\includegraphics[scale=.8,viewport=27 80 578 527]{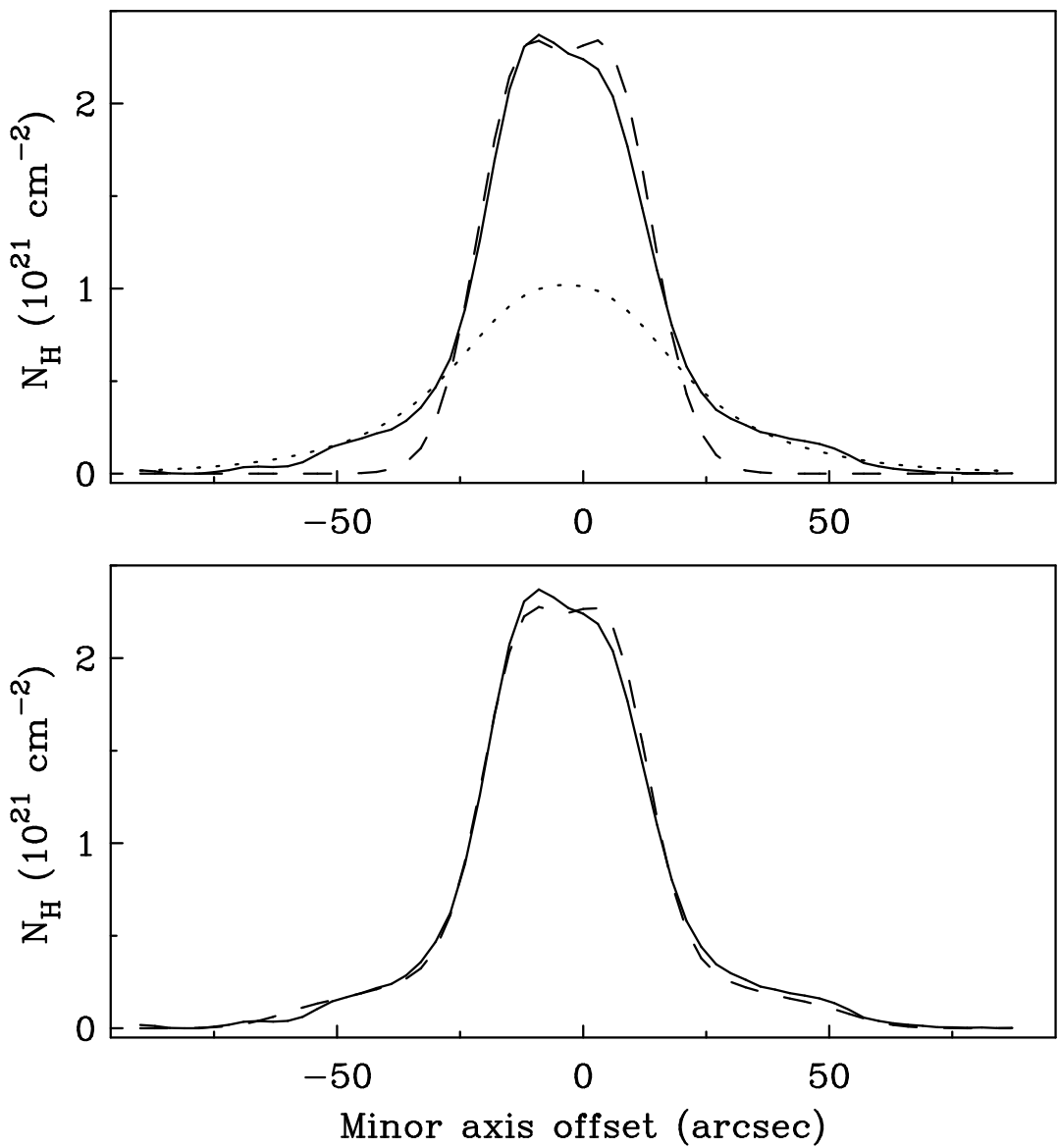} 
\caption{The solid line in both panels shows the minor axis emission profile
averaged over the inner 3' (25.6 kpc) of the major axis, in column density
units.  The dashed and dotted lines in the top panel show profiles for models
as described in the text with exponential scale heights of 3" and 18",
respectively.  The dashed line in the bottom panel shows the profile for Model
A, as described in the text.  The profiles for the other models described are
almost identical to that of Model A.
\label{fig9}} \end{figure}
\clearpage

Instead we consider two possibilities to explain the tails.  The first is a
warp along the line of sight, and the second is a halo with a box-like
(employed for ease of modeling although it is unphysical) profile in $z$.  The
model parameters are summarized in Table 1.

The morphology of the tails turns out to be very much coupled with their
kinematics, so we now consider these together by introducing pv diagrams
parallel to the minor axis at various positions along the major axis which
provide good leverage on the parameters of interest.  These are shown in
Figure 10, along with several models generated by GALMOD and variants on that
program.  The characteristic appearance of the disk component in these
diagrams is one of narrow angular extent at velocities furthest from $V_{sys}$
which broadens and (where the ringlike structure is evident) splits as
velocities move towards $V_{sys}$.  The faint high-latitude component has a
characteristic appearance which will greatly constrain its morphology and
kinematics.  Although it varies significantly among cuts, it generally
manifests itself as an extended, spectrally very narrow component at
velocities closer to $V_{sys}$ than the disk component, with a definite
gradient in $V_{hel}$ with $b$ that is more evident in panels
further from the galaxy center.  Some east-west asymmetries are
evident in this component.

\clearpage
\begin{figure} \epsscale{1.0}
\includegraphics[scale=1.0,viewport=17 121 578 697]{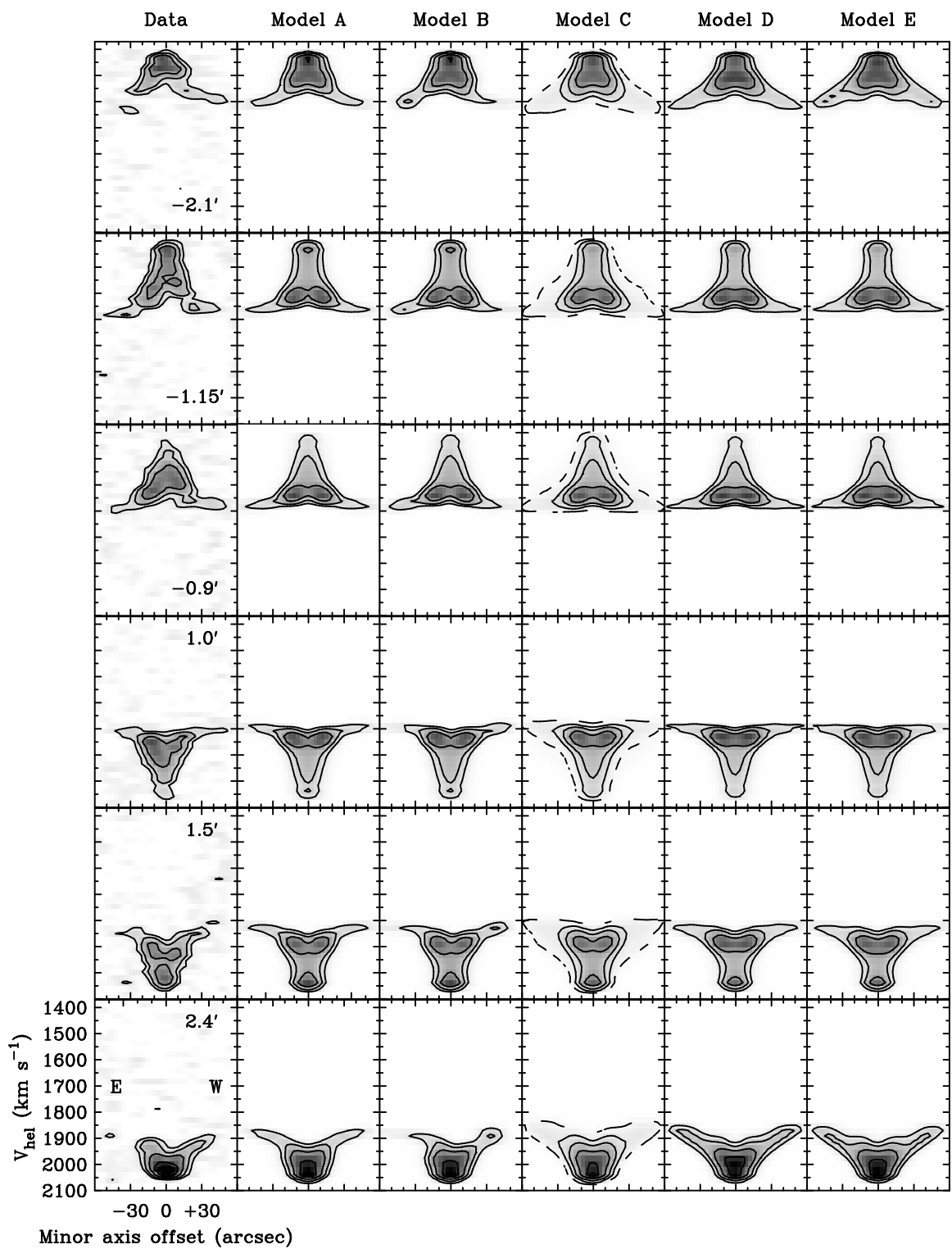} 
\caption{Position-velocity diagrams parallel to the minor axis at various
offsets along the major axis from the full-resolution data cube and various
models described in the text.  Contour levels are as in Figure 1.  The offsets 
along the major axis are shown in the panels for the data.
\label{fig10}} \end{figure}
\clearpage

We first consider a warp along the line of sight.  In GALMOD, this is achieved
by slowly decreasing the inclination of outer rings.  However, in order to
reproduce the minor axis emission profile, the column density in the warped
rings had to be increased significantly from the major-axis-based starting
point, as shown in Figure 6.  The warp begins at a radius of 4' (34.2 kpc),
where $i=84^\circ$, dropping to $79^\circ$ by 5' (42.8 kpc) radius (Figure
11).  The maximum displacement of the midplane of the outermost warped ring
from the unwarped disk is about 5 kpc.  A slightly larger warping is required
for the SE quadrant (top three panels of Figure 10 at negative minor axis
offsets) but is not modeled here.  The resulting model, Model A, appears to
match the observed minor-axis pv diagrams very well.
\clearpage
\begin{figure} \epsscale{1.0}
\includegraphics[scale=1.0,viewport=7 41 578 547]{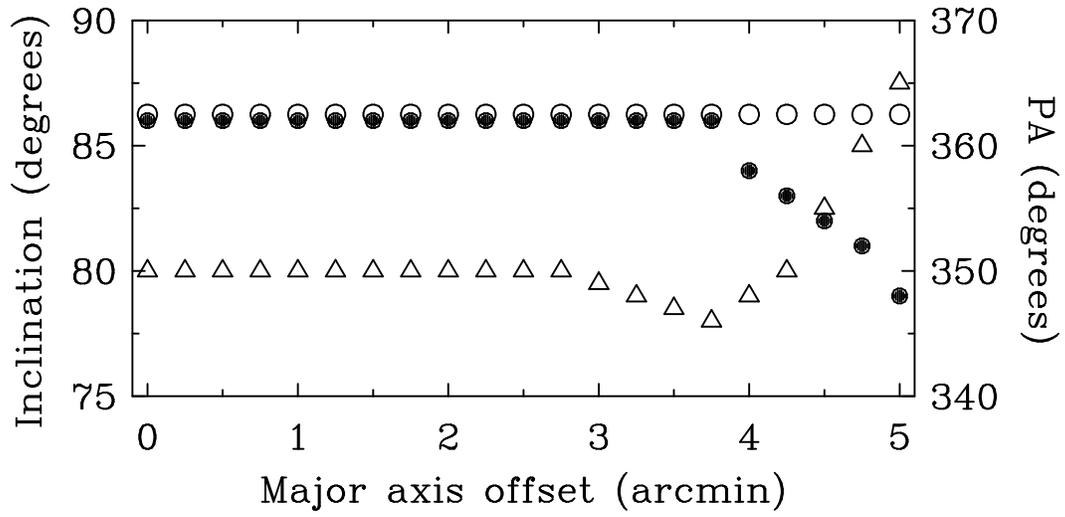} 
\caption{Filled circles show the run of inclination (left scale) for
Models A and B.  Open triangles show the run of PA (right scale) for Model B.
Open circles show the inclination for Models C, D and E.
\label{fig11}} \end{figure}
\clearpage

Model B is an attempt to incorporate the bending across the line of sight
evident in the zeroth-moment maps.  Retaining the run of inclination in Model
A, the PA in this model decreases from 350$^\circ$ to 346$^\circ$ over the
radius range 3' to 4', and then increases back to 5$^\circ$ by 5' radius
(Figure 11).  The asymmetry introduced into the minor-axis pv diagrams (Figure
10) provides a slightly better match to the data at high latitudes, although
still probably underestimating the observed asymmetry.  The zeroth-moment map
of this model (Figure 12) also exhibits a slight brightness asymmetry with
respect to the minor axis at major axis distances less than 4' from the center
and at heights 30--60'' from the plane that is also apparent on the east side
in Figure 2.  The model appears to somewhat overestimate the bending to larger
PAs evident in Figure 2 (and, of course, there is too much emission at the
ends of the major axis because of the excess column density in the modeled
warp).  However, a model version smoothed to 60'' (Figure 12, center), is a
reasonable match to the bending in Figure 4.  We note that warps are often
asymmetric \citep{2002A&A...394..769G}, so one should not expect a warp fitted
to kinematic information parallel to the minor axis to match perfectly the
morphology along the major axis.  A second caveat to this exercise is that
some of the bending along the major axis may be due to outer spiral structure
seen not quite edge-on.  In fact, the initial bending to lower PAs in the
model is only present to match the morphology in the moment maps.  The
subsequent increase in PA is responsible for the asymmetries in the pv
diagrams.  As mentioned at the beginning of this section, for these kinds of
reasons, we have not carried out a full exploration of parameter space but
simply present these models as reasonable fits to the data.

\clearpage
\begin{figure} \epsscale{.80}
\includegraphics[scale=.8,viewport=8 159 480 763]{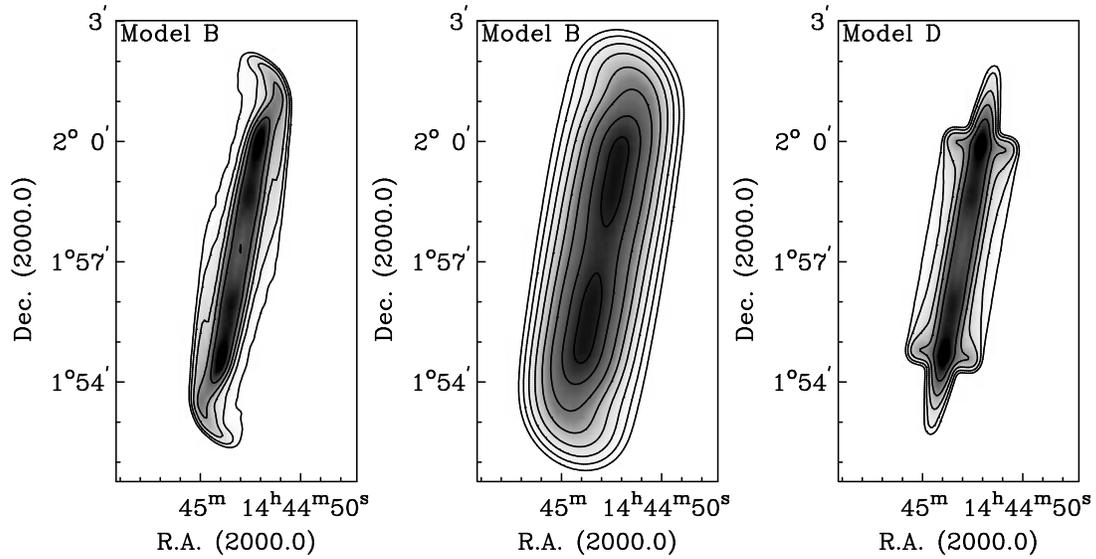}
\caption{{\it Left:} Zeroth-moment map of Model B, convolved to 15'' resolution.
Contour levels are as in Figure 2 (left).
{\it Center:} Zeroth-moment map of
Model B, convolved to 60'' resolution.   Contour levels are as in Figure 3.
{\it Right:} Zeroth-moment map of Model D, convolved to 15'' resolution.
Contour levels are as in Figure 2 (left).
\label{fig12}}
\end{figure}
\clearpage

We next consider the box-like halo, extending to $z=50"$ (7.1 kpc).  Such a
component, when convolved with the beam, loses its sharp edges and, added in
the right proportion to the disk component, reasonably reproduces the
minor-axis emission profile in Figure 9, being almost indistinguishable from
the warp model profile.  The disk component column density is normalized
downwards by a factor of 1.24 in these models.

For the halo models, the gradient in $V_{hel}$ with $b$ at high latitudes in
Figure 10 implies a rotational velocity that decreases with $z$,
i.e. $dV_{rot}/dz < 0$.  The spectral narrowness implies that such a halo
cannot occupy a large range in radius, otherwise projection of the rotational
velocity would cause broadening of the $V_{hel}$ profiles in disagreement with
the data.

To illustrate these constraints, we first consider a model where the halo has
the same radial density profile shape as the disk, and a vertical gradient in
rotation velocity of --3.5 km s$^{-1}$ (arcsec)$^{-1}$ (--25 km s$^{-1}$
kpc$^{-1}$). We refer to this as Model C in Figure 10.  Although the amount of
high-latitude emission is well matched to the data, there is little such
emission seen at the first solid contour level of this model.  This is because
the velocity profiles there are broader with lower peak intensities, extending
to velocities much too far from $V_{sys}$ to be consistent with the data, and
we demonstrate this by including a low-level dashed contour. Clearly, this
model is a poor match to the halo emission.

The narrowness of the high-latitude velocity profiles puts a strong constraint
on the radial range of the halo component, and leads us to consider annular
distributions, beginning with one in which the entire halo is contained in one
15" ring.  Model D in Figure 10 features a ring of central radius $R_0$ =
180", with $dV_{rot}/dz = -3.5$ km s$^{-1}$ arcsec$^{-1}$, and a necessarily
high column density of $9 \times 10^{20}$ cm$^{-2}$ to match the minor axis
emission profile.  Comparison of modeled and data pv diagrams for other values
of $R_0$ constrain this parameter to be in the range 160--200" (22.8--28.5
kpc).  Even for such a narrow ring, the high-latitude velocity profiles tend
to be broader than in the data.  For values of $dV_{rot}/dz$ outside the range
--3 to --4 km s$^{-1}$ arcsec$^{-1}$ the gradient in the halo component in the
pv diagrams is not well matched.  However, a somewhat higher gradient is
required for the SE quadrant, and models indicate a value of $dV_{rot}/dz =
-5$ km s$^{-1}$ arcsec$^{-1}$.  All halo models employ $i=86.25^\circ$ to
match the minor-axis emission profile.

The narrowness of the ring also introduces a strong edge-brightening in the
halo in zeroth-moment maps (Figure 12) due to the projection of the ring which
is clearly not seen in the data.  This halo would therefore need to be much
more asymmetric than the warp of Models A and B.  The normalization of the
modeled halo also now depends on the radial range used to form the minor axis
emission profiles.  We have chosen to normalize using the same radial range as
above, realizing that the edge-brightening will lead to excess mass in the
modeled halo.

An annulus wider than 1' (8.6 kpc), although featuring a more plausible column
density, causes the high-latitude velocity profiles to broaden, if only
slightly.  Model E, where the annulus is 1.25' wide and its column density is
$1.8 \times 10^{20}$ cm$^{-2}$, shows this effect.

The major-axis pv diagram for the data and Models B and D are compared in
Figure 7.  The general shape of the flat part of the pv diagram is reproduced
quite well, given that we do not attempt to model the aforementioned asymmetry
in the emission around the major axis.  The general slope of the rising
part of the diagram is underestimated.  In the models, the slope is
dictated by the fact that most of the gas is in the outer ($R=3'$) ring
in our axisymmetric radial profile.  Asymmetries, in the sense of gas
being more centrally concentrated in the vicinity of the inner ring
at major axis offsets within 1' of the center, could be the cause
of this discrepancy.

The velocity dispersion is not well constrained because of the somewhat
low velocity resolution.  A value of 10 km s$^{-1}$ was used in the models
for both disk and halo.

We roughly estimate the fraction of mass in the extended component in two
different ways.  First, we simply assume that in the zeroth-moment map of
Figure 2, all emission above a certain height can be attributed to the
extended component.  Using two values of this height -- 9" and 11" (1.3 and
1.6 kpc) --, its mass is $1.2-1.6 \times 10^9$ \msun.  The second estimate
comes from the models.  In the warp model, the excess column density in the
warped region accounts for $1.8 \times 10^9$ \msun.  For the halo models, we
calculate the mass in the halo component and find $3.1 \times 10^9$ \msun.
These are both overestimates, especially the latter, because the extra column
density along the line of sight required by the axisymmetric models
overestimates the emission along the major axis.  The estimate from the data
themselves is likely to be superior.

Although both models are inaccurate because of their symmetry, the problem is
much worse for the halo model.  This model is also much more contrived in that
it requires a very large mass in a very small radial range.  The narrowness of
the high-latitude spectra is also better reproduced by the warp model.  We
therefore conclude that the extended component is a warp rather than a halo.
Further fine tuning of the warp model may produce a slightly better match to
the data, but the asymmetries preclude a significant improvement, and hence we
do not pursue this further here.

\subsection{Radial Inflow}

Figure 13 (top panel) shows a pv diagram along the minor axis of the galaxy.
A shift in mean velocity between the positive and negative sides is apparent
in the disk component, while a very small shift also exists in the warp
component (the double-peaked nature of the disk component is due to the inner
ring of 1.5' radius as discussed above).  Such a shift is a potential sign of
radial inflow in a not quite edge-on galaxy.  All diagrams for major axis
offsets within one beam width of this one show this signature.  For the warp,
such a shift is also seen, albeit at a slightly smaller amplitude, in the
corresponding pv diagram of Model B (bottom panel of Figure 13), and we cannot
rule out that it is an effect of the warp geometry.  For the disk, since there
is strong kinematic evidence of a bar (see \S 1), the signature would suggest
radial inflow in the inner ring-like gas distribution surrounding the bar.  To
measure the inflow amplitude, we have created two spectra from the minor axis
pv diagram, averaged over the range 3--18" in the minor axis direction on the
east and west sides.  A Gaussian is fit to each spectrum to find the mean
velocity, and half of the velocity difference between the east and west sides
is taken as the inflow velocity.  The result is $V_{inflow} = 9.7 \pm 0.4$ km
s$^{-1}$.
\clearpage
\begin{figure} \epsscale{.80}
\includegraphics[scale=.8,viewport=0 91 578 797]{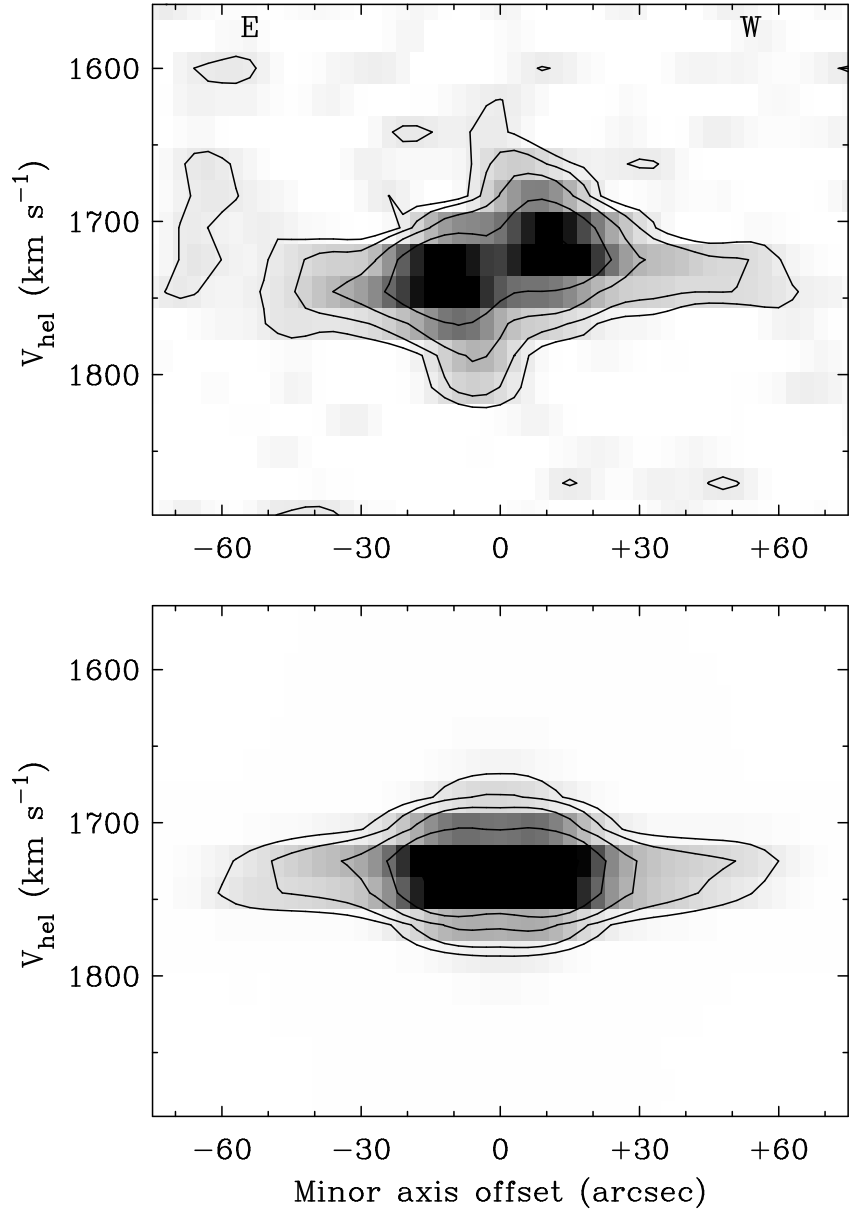} 
\caption{Minor-axis position-velocity diagram for the full-resolution
cube (top) and Model B (bottom).  Contour levels are 
1.5, 3, 6 and 12 times the 1 $\sigma$ noise in column
density units of $4.6 \times 10^{19}$ cm$^{-2}$.
\label{fig13}} \end{figure}
\clearpage
\subsection{A Possible Supershell}

A structure in the SW quadrant above the disk has the form of a closed shell
-- better seen in Figure 14 -- although it is always possible that projection
of unrelated features has given it this appearance.  The total mass is about
$10^8$ \msun, and the extent parallel to the major axis is about 3.4 kpc.
Position-velocity diagrams parallel to the minor axis for the northern wall,
center and southern wall of the shell are shown in Figure 15.  Above about 10"
(1.4 kpc) from the plane, the walls show emission at the velocities of the
warp as well as a broad component extending to about the terminal velocity in
the northern wall, and some evidence for line splitting in the southern wall.
The northern wall is detected over nine velocity channels, or about 190 km
s$^{-1}$.  The line splitting in the southern wall is five channels or about
100 km s$^{-1}$.  Given that the velocities in the warp component are due to
the warp geometry, it would be incorrect to conclude that this line
broadening/splitting is due to expansion.  Rather, it may just indicate that
the feature is extended along the line of sight.  There is no indication of a
corresponding ionized feature in the H$\alpha$ image of
\citet{1996ApJ...462..712R}.

\clearpage
\begin{figure} \epsscale{1.0}
\includegraphics[scale=1.0,viewport=100 141 578 697]{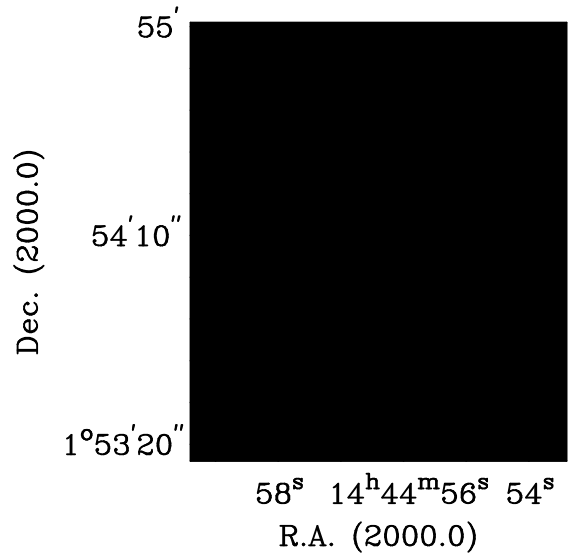} 
\caption{Close-up of the zeroth-moment map of NGC 5746 showing the
shell-like structure described in the text.
\label{fig14}} \end{figure}

\begin{figure} \epsscale{1.0}
\includegraphics[scale=1.0,viewport=-100 141 578 597]{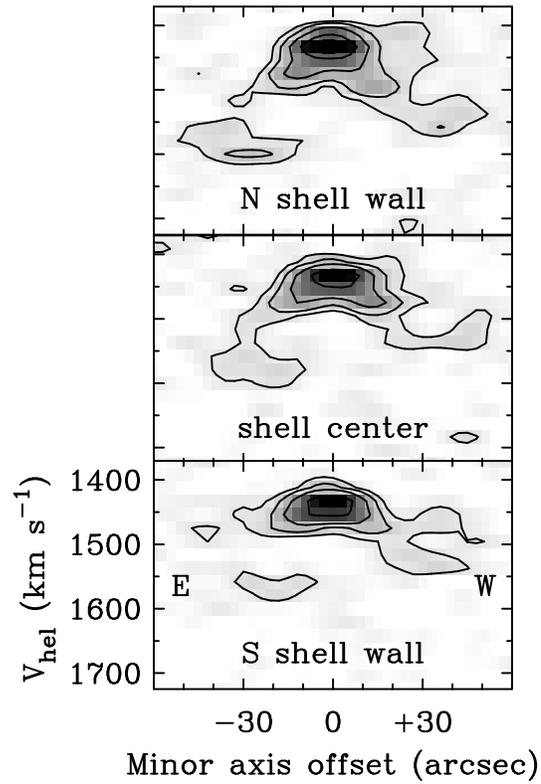} 
\caption{Position-velocity diagrams parallel to the minor axis from the
full-resolution cube for the northern wall, center and southern wall of the
possible supershell.  Contour levels are 2, 4, 8, 16 and 24 times the 1
$\sigma$ noise in column density units of $4.6 \times 10^{19}$ cm$^{-2}$as in
Figure 1.
\label{fig15}} \end{figure}
\clearpage

\subsection{A Clumpy Warp?}

Figure 2 shows that the high-latitude emission is partially resolved into
individual clumps.  Almost thirty such features have been identified by visual
inspection and are listed in Table 2.  This selection is not meant to be
complete or statistically well defined, but is simply meant to give a
first-order idea of the masses of well detected features.  These range from
about $4 \times 10^6$ \msun\ to $10^8$ \msun, and their masses sum to $7
\times 10^8$ \msun, or about half of the total mass of the high-latitude
emission.  The most massive, numbered 18, is the shell-like structure
discussed above.  Most are unresolved but a few have a filamentary appearance
and lengths of 3--6 kpc.  The line widths are difficult to estimate for many
features because of the velocity resolution of 42 km s$^{-1}$ (FWHM) and the
low signal-to-noise ratio of some of the detections.  For example, the
significant emission for three of the best detected clouds, numbered 5, 6 and
8 in Table 2, is confined to two channels and therefore the lines are
unresolved.  Others, such as 11 and 28, show emission over $6-7$ channels, or
$125-145$ km s$^{-1}$, and it is not clear if these represent actual internal
motions or reflect the extent of the features along the line of sight, for
example.

The 60"-resolution zeroth-moment map (Figure 4) reveals two faint vertical
extensions from the plane in the SE quadrant, reaching heights of 15 and 20
kpc.  Summing their emission above where they appear to merge with the disk,
we obtain masses of $7.9 \times 10^6$ \msun\ and $6.8 \times 10^6$ \msun, for
the northern and southern feature, respectively.  A faint extension is also
seen off the NW end of the disk.  Its mass is about $6.2 \times 10^6$ \msun.
The first-moment map shows that the SE features have about the same velocity
as the part of the warp that they connect to with little change in velocity
along their vertical extent.

\subsection{High Latitude Features Not in the Warp}

We have carefully combed the channel maps and pv diagrams to look for
high-latitude emission which does not have the velocities of the warp
component but which may not be evident in Figure 2.  We have found four such
features.  Position-velocity diagrams for the first three, at the center of
their extent along the major axis, are shown in Figure 16.  The features'
locations are also marked in Figure 3, and their properties are summarized in
Table 3.  One is an unresolved feature in the NE quadrant of the halo.  The
second and third are diffuse features with appreciable extents along the major
axis of 0.7' and 0.6' (6.0 and 5.1 kpc), respectively.  The second feature is
also counterrotating and its emission spans more than 300 km s$^{-1}$.  The
fourth is the low-$V_{hel}$ side of the possible shell (Cloud 18 in Table 2).
\clearpage
\begin{figure} \epsscale{1.0}
\includegraphics[scale=1.0,viewport=-100 121 578 697]{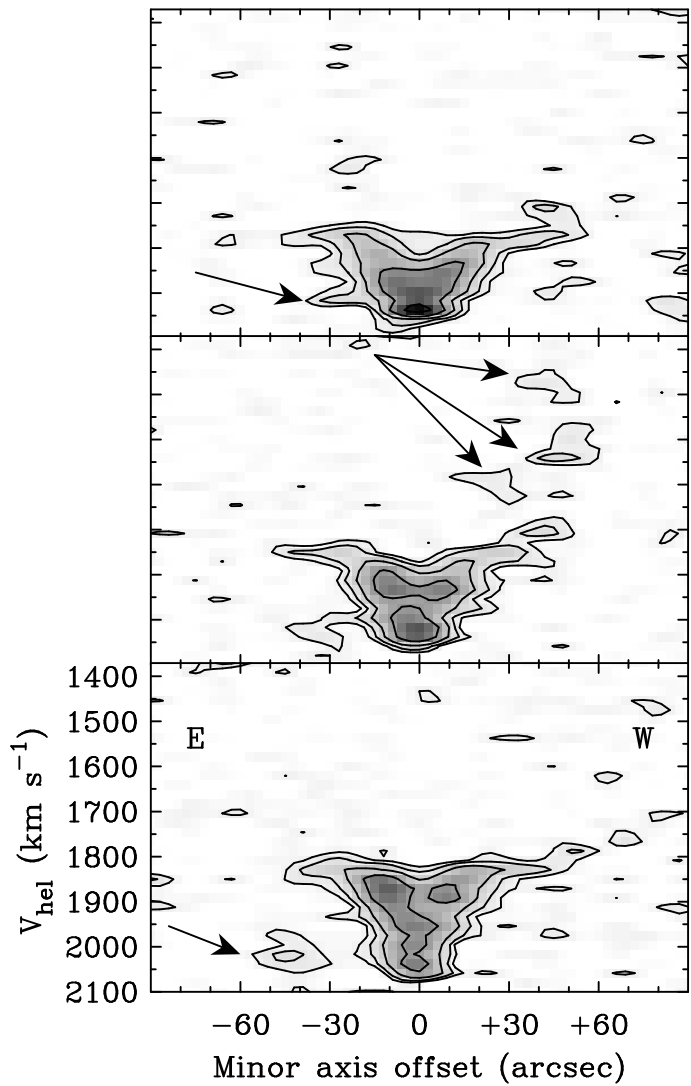} 
\caption{Position-velocity diagrams parallel to the minor axis from the full-resolution
cube showing emission (indicated by arrows) from three high-latitude features
that do not have the velocities of the warp.  From top to bottom, these are
Clouds 1, 2, and 3 in Table 3.  Contour levels are 1.5, 3, 6, 12, 24, 36
and 48 times the 1 $\sigma$ noise in column density units of 
$4.6 \times 10^{19}$ cm$^{-2}$.
\label{fig16}} \end{figure}
\clearpage

\subsection{NGC 5740}

As this barred \citep{2005MNRAS.364..283E} Sb galaxy is beyond the half-power
radius of the primary beam, the detailed structure may not be accurate and we
limit ourselves to general properties only.  We assume its distance to be the
same as that of NGC 5746.  Channel maps of the full-resolution cube are shown
in Figure 17.  Zeroth- and first-moment maps of NGC 5740 are presented in
Figure 18, using all data above $2\sigma$ in two consecutive channels.  The
zeroth-moment map is overlaid on a Digitized Sky Survey red image in Figure
19.  There is a disk of about 30 kpc diameter and a large extension to the
NNW.  The disk has a mass of about $5.3 \times 10^9$ \msun\ and the extension
$9.4 \times 10^8$ \msun.  For comparison, \citet{1986AJ.....92..742S} quote a
mass in atomic hydrogen of $7.0 \times 10^9$ \msun, scaled to our distance.
\clearpage
\begin{figure} \epsscale{1.0}
\includegraphics[scale=1.0,viewport=17 121 578 697]{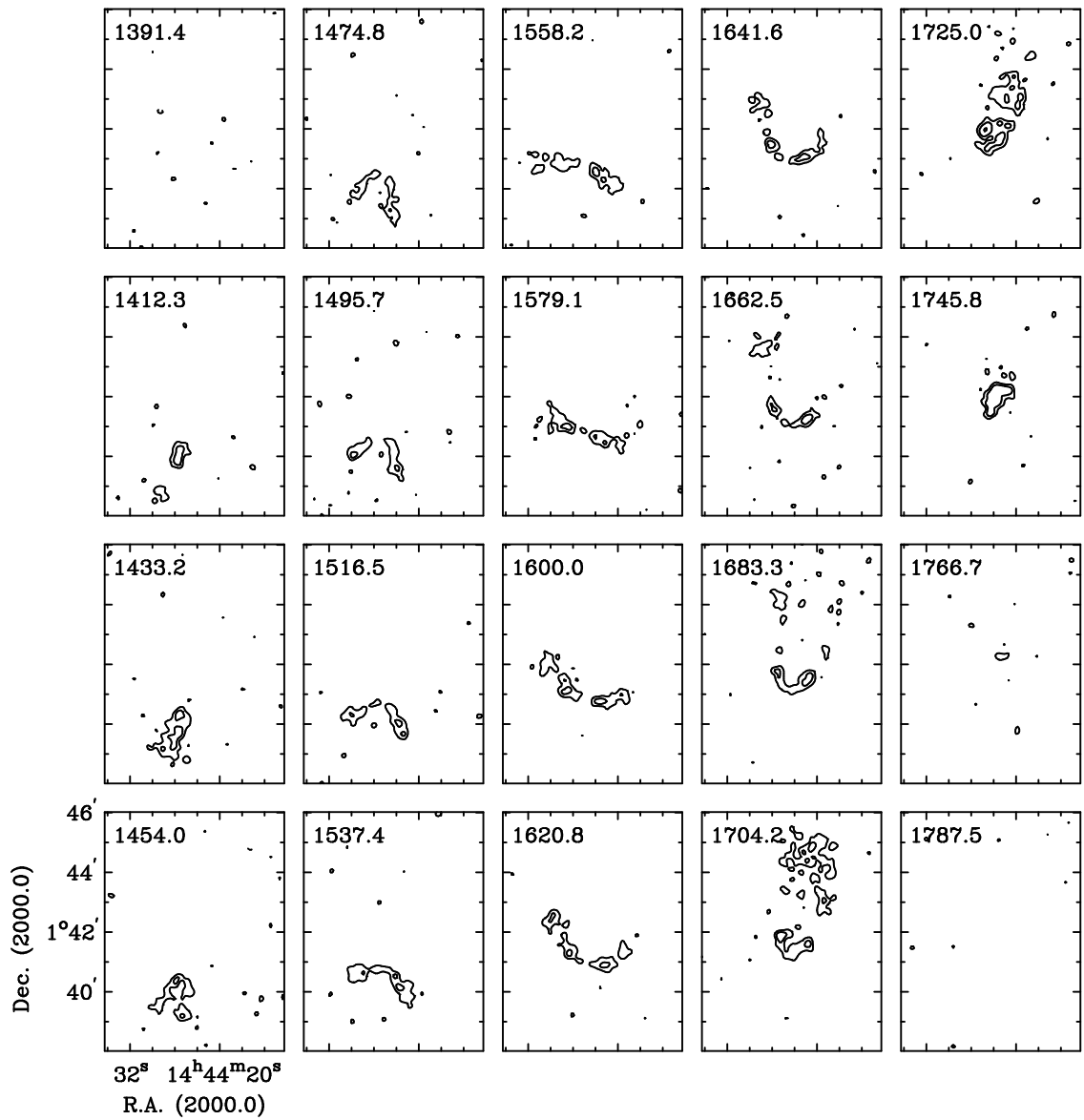}
\caption{Channel maps from the full-resolution cube for NGC 5740, corrected
for primary beam attenuation.  Contour levels are 3, 6, 12, 24, 36 and 48
times $4.6 \times 10^{19}$ cm$^{-2}$.  Each map is labeled with its
heliocentric velocity.
\label{fig17}}
\end{figure}

\begin{figure} \epsscale{1.0}
\includegraphics[scale=1.0,viewport=67 121 578 697]{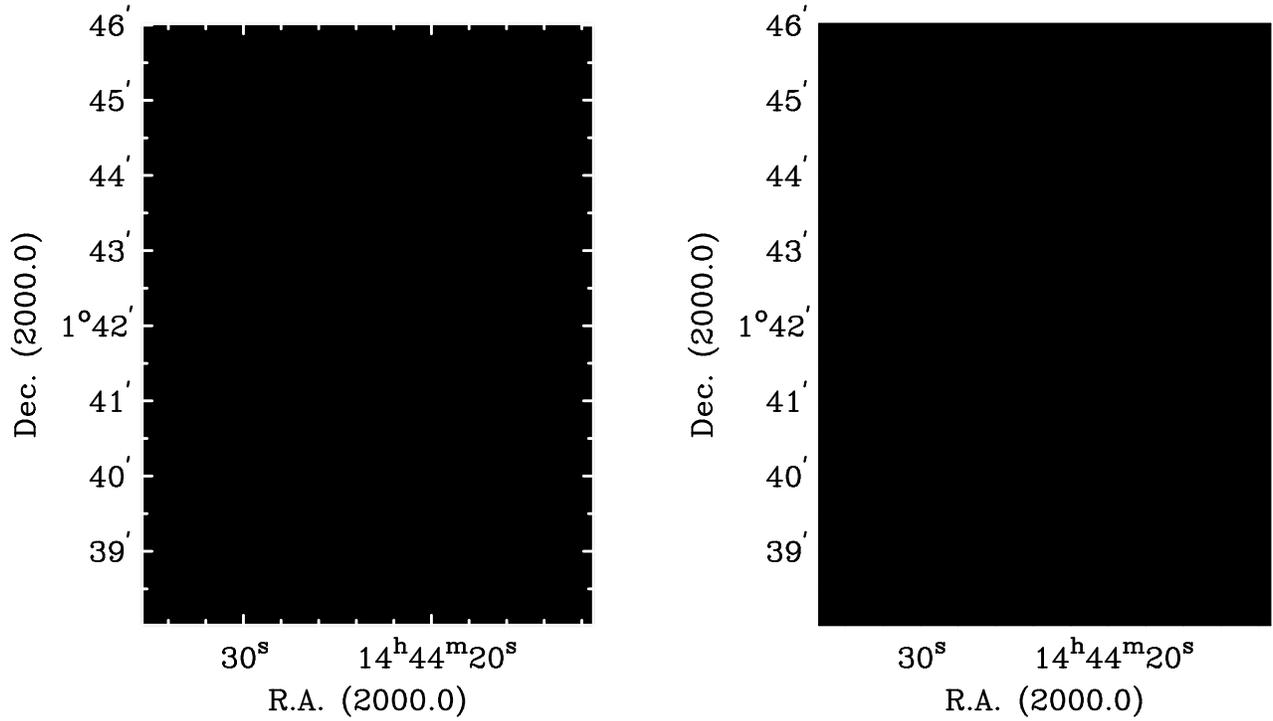}
\caption{
{\it Left:} Zeroth-moment map of NGC 5740 from the full-resolution,
primary-beam-corrected cube.  Contour levels are 2, 4, 8, 12 and 16 times
$10^{20}$ cm$^{-2}$. {\it Right:}  First-moment map of NGC 5740. Contours run
from 1440 to 1740 (the closed contour on the NE side at Decl. 1$^\circ$ 42') 
km s$^{-1}$ in 20 km s$^{-1}$ increments from SW to NE.
\label{fig18}}
\end{figure}

\begin{figure}
\epsscale{0.6}
\includegraphics[scale=1.0,viewport=17 21 578 637]{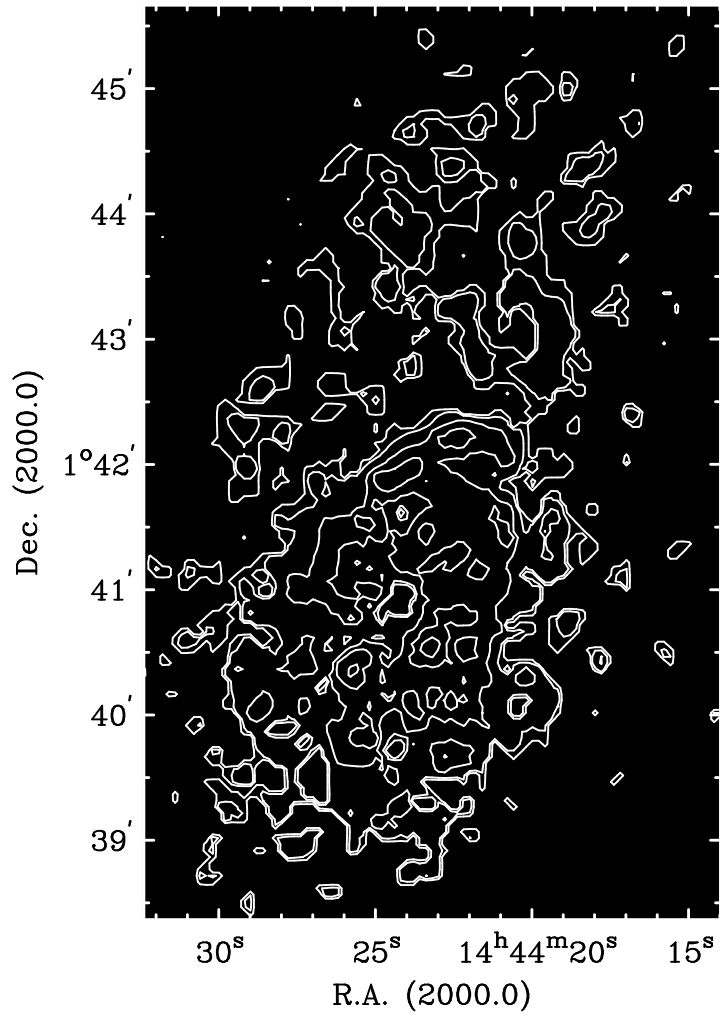}
\caption{Zeroth-moment map of NGC 5740 from the full-resolution
cube overlaid on a Digitized Sky Survey red image.  Contour
levels are as in Figure 18 (left).
\label{fig19}}
\end{figure}
\clearpage

A ROTCUR analysis of the disk indicates a PA of $339 \pm 0.5^\circ$, an
inclination of $58 \pm 2^\circ$, $V_{sys} = 1586 \pm 6$ km s$^{-1}$, a
dynamical center of R.A. 14$^{\rm h}$ 44$^{\rm m}$ 24.4$^{\rm s}$,
Decl. 1$^\circ$ 40' 52", with approximate uncertainty 3" in R.A. and 8" in
Decl.  The first-moment map suggests regular rotation continues into the NNW
extension.  Keeping $V_{sys}$ and the dynamical center fixed, and fitting the
receding half of the galaxy only, the extension exhibits little change in PA
but a continued fall in $V_{rot}$ and an increase in inclination to $61^\circ$
at 2' and $70^\circ$ at 4'.  The derived rotation curve is shown in Figure 20.
The dynamical mass within $R=2'$ (17.1 kpc) is $1.2 \times 10^{11}$ \msun.

Morphologically, the NNW extension is similar to extensions in several Virgo
Cluster galaxies, most of which are best explained by ram pressure stripping
\citep{2007ApJ...659L.115C}.  There is no obvious evidence of compression of
the gas on the opposite side of the disk, however.  The condition for ram
pressure stripping, as given by \citet{2001ApJ...561..708V}, is
$\rho_{IGM}\,v^2_{gal} > \Sigma_{ISM}\,V_{rot}^2/R$, where $\rho_{IGM}$ is the
intragroup medium (IGM) density, and $v_{gal}$ is the velocity of the galaxy
through the medium, for gas at surface density $\Sigma_{ISM}$ at a radius $R$
from the galactic center.  At the outskirts of the main disk the typical
surface density is about 2 \msun pc$^{-2}$, while $V_{rot} = 175 $ km
s$^{-1}$.  The condition for stripping is then ($n/10^{-3}$ cm$^{-3}$)($v$/100
km s$^{-1}$)$^2> 14$.  Both required numbers are not constrained.  We find
that the one-dimensional velocity dispersion for the catalogued galaxies of
the NGC 5746 group is 120 km s$^{-1}$.  The IGM density is unknown.  X-ray
bright groups are found to be HI-deficient by \citet{2006MNRAS.369..360S}, who
calculate a range of IGM densities from $5 \times 10^{-4}$ to $2 \times
10^{-3}$ cm$^{-3}$, but it is unknown how extensive X-ray emission is in the
NGC 5746 group.  Hence, it is unclear whether the stripping condition can be
met.  Tidal stripping seems less likely given the morphology and that the
closest group galaxy on the sky, NGC 5746, is 155 kpc away in projection.

\clearpage
\begin{figure} \epsscale{.80}
\includegraphics[scale=.8,viewport=47 41 578 547]{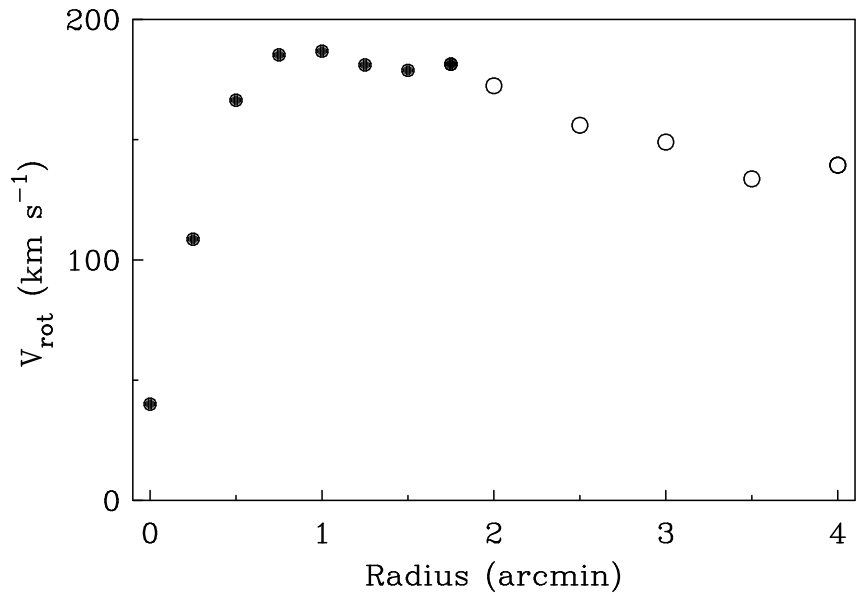} 
\caption{Filled circles show the rotation curve for NGC 5740 derived from
a tilted ring analysis.  Open circles show the extension of the
rotation curve into the NNW extended emission component.
\label{fig20}} \end{figure}
\clearpage

\section{Discussion}

Almost all of the high-latitude gas we have detected can be associated with
the warp.  At the level of sensitivity of our data, only the features listed
in Table 3, summing to about 10$^8$ \msun, may have originated in a disk-halo
flow or infall.  The heights off the plane of their centers range from 2 to 8
kpc.  One of the more massive ones is counter-rotating and is unlikely to have
originated in the disk.  A few$\,\times\, 10^7$ \msun\ worth of
counter-rotating clouds were also discovered in the neutral halo of NGC 891 by
\citet{2007AJ....134.1019O}.

Are these few clouds consistent with predictions from the halo thermal
instability simulations mentioned in \S 1?  In the simulations of
\citet{2006ApJ...644L...1S}, most of the warm clouds are within 50 kpc from
the center of the galaxy and total $10^8$ \msun, with most having masses a few
times $10^5$ to a few times $10^6$ \msun\ for a galaxy like the Milky Way.  By
comparison, for an M33-like galaxy, \citet{2006MNRAS.370.1612K} expect clouds
of radii $0.1-0.6$ kpc and masses from $10^5$ to a few times $10^6$ \msun,
totaling $\sim 10^7-10^8$ \msun\ depending on the simulation parameters,
confined to within about 10--20 kpc above the disk.  If the few clouds in
Table 3 are due to infall, we note that their total mass is comparable to
these model predictions, with the caveat that the fraction of neutral gas is
uncertain in the simulations.  The small number of detected clouds is in fact
comparable to that in the ``very high resolution'' simulation of
\citet{2006ApJ...644L...1S}, where the mass resolution is comparable to that
of our data.

For galaxies like the Milky Way, \citet{2004MNRAS.355..694M} predict $2 \times
10^{10}$ \msun\ of mostly ionized clouds of mass $5 \times 10^{6}$ \msun\ and
size $\sim 1$ kpc, extending to 150 kpc from the center of the galaxy.  At our
sensitivity, it is difficult to rule out with confidence such a widespread
population of clouds from the HI data alone.

It would be interesting to know how much mass in warm clouds forms in
simulations of more massive systems like NGC 5746, and where it is located.
The cooling rate, at least, of the hot gas in this galaxy can be estimated
from the observations, but is very uncertain.  The rate inferred by
\citet{2006astro.ph.10893R}, who exclude the projected disk area where there
may be contamination by other sources of hot gas, is only 0.2 \msun\
(yr)$^{-1}$.  The rate is modest because the estimated halo temperature is
rather high, at 0.56 keV.  However, most of the halo cooling is likely to be
occurring in the excluded area, and their models of massive disks suggest that
the cooling rate could be 5--10 times higher as a result.  In summary, it is
unclear whether the observed clouds are consistent with theoretical
expectations for halo cooling in a massive disk galaxy like NGC 5746.

What fraction of these halo clouds do we expect to be fully or mostly ionized?
This depends on the incident ionizing flux, the mass, and most importantly,
the bounding pressure on the cloud. If we assume that the ionized gas in a
cloud is in thermal pressure equilibrium with an outside pressure, the density
of a cloud is given by $n_{-2}=4.55 T_{4}^{-1}\left[p_{3}/k\right]$, where
$n_{-2}=n_{H}/10^{-2}~{\rm cm^{-3}}$ is the normalized hydrogen density of the
cloud, $T_{4}=T/(10^{4}~{\rm K})$ is the normalized temperature, and $\left[
p_{3}/k \right]=[p/k]/(10^{3}\, {\rm cm^{-3}~K})$ is the normalized thermal
pressure.  (We use here $\rho/n_{H}=2.4 \times 10^{-24}$ for solar metallicity
gas, $n_{tot}=2.2n_{H}$, and $n_{e}=1.2n_{H}$ for fully ionized plasma). For a
constant density spherical cloud, the cloud radius is given by $R=(532~ {\rm
pc}) M_{6}^{1/3} T_{4}^{1/3} \left[ p_{3}/k \right] ^{-1/3}$.

We assume that the outer skin of this cloud will be ionized by a
(plane-parallel) flux of H ionizing photons, $\phi_{5}=\phi / (10^{5}~{\rm
photons~ cm^{-2}\, s^{-1}})$. The metagalactic flux is estimated to be $\phi_{5}
< 0.45$ (\citealt{2002ApJS..143..419S}, and references therein), while high
velocity clouds in the Galactic halo experience $\phi_{5}=2$
\citep{2002ApJ...572L.153T}. The thickness of this skin, $\ell$, is determined
by the balance of ionizations and recombinations,
$\phi=\alpha_{B}n_{e}^{2}\ell$ where the Case B recombination coefficient is
$\alpha_{B} \cong 3 \times 10^{-13}\, {\rm cm^{3}\, s^{-1}}$. (Note that for
optically thick clouds, $\phi=\alpha_{B}~{\rm EM}$, so that the surface averaged
emission measure is a simple measure of the ionizing flux.)  In this case, the
skin depth is given by $\ell=(35.7~ {\rm pc}) \phi_{5} T_{4}^{2} \left[
p_{3}/k \right] ^{-2}$.

With these approximations, the ionized volume and mass of the cloud is given
by $f_{ion}=1-[1-(\ell/R)]^{3}$ or $f_{ion}=3(\ell/R)-3(\ell/R)^{2}+...$ for
small $(\ell/R)$. The condition for a cloud to be fully ionized is

\begin{equation} 
 \left[ \frac{p}{k} \right] < (197~ {\rm cm^{-3}~K})~ \phi_{5}^{3/5}T_{4}M_{6}^{-1/3} . 
\end{equation} 

For very low ionization fraction ($f_{ion} < 5$\%), this criterion is well
approximated by

\begin{equation}
  \left[ \frac{p}{k} \right] = (380~{\rm cm^{-3}~K}) f_{ion}^{-3/5}  \phi_{5}^{3/5}T_{4}M_{6}^{-1/3}.
\end{equation}

\noindent so that an ionization fraction of less than 1\% requires  $[p/k] > (6022~ {\rm cm^{-3}~K})~ \phi_{5}^{3/5}T_{4}M_{6}^{-1/3}. $  

The sensitivity of the ionization fraction of gas clouds to the outside
pressure has been studied by \citet{1994ApJ...423..665F},
\citet{1995ApJ...453..673W}, and \citet{2003ApJ...589..270M} with comparable
results. The normalization constant relating the pressure to the ionization
fraction and cloud mass will depend on the assumed geometry.  The presence of
central condensation due to dark matter halos can also lower somewhat the
pressure necessary to form neutral clouds \citep{2002ApJS..143..419S}.

These considerations show that not only are X-ray halos associated with galaxy
formation more likely to form condensations, but that the higher pressures of
such halos result in a much larger neutral fraction for the
condensation. Using the density and temperature estimate for the X-ray halo of
NGC 5746 \citep{2006astro.ph.10893R}, the halo pressure is given by
$[p/k]=6200~ {\rm cm^{-3}~K}$. Such a high bounding pressure would guarantee
that clouds embedded in this hot gas will be primarily neutral.  It is also
interesting to note that the bounding pressure for the cloud can be either the
thermal gaseous halo pressure or the ram pressure associated with a cloud's
motion through the halo. The one cloud showing counter rotation presumably has
a much higher ram pressure than most of the rest of neutral gas detected and
could be expected to have a higher neutral fraction.

At larger radius than that traced by the X-ray gas, it remains possible that
the pressure will drop to such a point that a large population of fully
ionized clouds might be present, but uncertainties in the X-ray data and
analysis prevent such pressure variations from being assessed.  Although the
predicted emission measure from the front and back face of an individual cloud
is a relatively low EM$= 2 \times (0.11~{\rm pc~cm^{-6}}) ~\phi_{5}$, a
population of unresolved smaller clouds in the beam might boost the H$\alpha$
emission to detectable levels. However, processes such as thermal conduction
and evaporation should set a lower limit on the mass of cloud that can survive
\citep{2004MNRAS.355..694M}.  We have reexamined the H$\alpha$ image of
\citet{1996ApJ...462..712R}, smoothed to 1.5'' resolution -- yielding a noise
level of 2.2 pc cm$^{-6}$ -- and find no evidence for such bright clouds in
the halo.

To attempt to constrain the origin of the neutral gas halo in this and other
galaxies, we can compare the neutral halo mass and mass fraction with
estimates for other galaxies and look for correlations with other parameters
(Table 4).  In this comparison one should keep in mind that some of the
galaxies are edge-on, one (NGC 6946) very much face-on, and some at
intermediate inclinations (NGC 4559, NGC 2403 and NGC 253), and thus the
characterization of halo gas necessarily varies from galaxy to galaxy.  In the
galaxies not viewed edge-on, the extraplanar gas is identified by its
anomalous velocities compared to the bulk of the emission.  We also note that
the sensitivity and linear resolution of these observations vary
significantly.

NGC 5746 ranks near the bottom of the table in terms of halo mass and mass
fraction and also has one of the lowest levels of star formation activity, as
estimated from $L_{FIR}$/$D_{25}^2$, lending support for an origin in a weak
disk-halo flow.  However, there is no discernable trend among the galaxies in
this direction.  NGC 6946 and NGC 253 stand out as having little HI in their
halos given their star forming activity.  The halo HI in the latter is
particularly puzzling as it is found only on one side of the disk, away from
the center of the galaxy \citep{2005A&A...431...65B}.  It is possible that the
neutral gas has been swept radially outward by the central starburst outflow
or that much of the halo gas is ionized.  In NGC 6946,
\citet{2007PhDT.........1B} note that the mass of halo gas more than doubles
in a map at 64'' resolution vs.  one at 22'' resolution.  The rotation speeds
of the galaxies span a factor of three, and thus the ability of supernova
power to raise gas off the plane should vary significantly from galaxy to
galaxy.  In particular, the high fractions of halo HI in NGC 4559, NGC 2403
and UGC 7321 could be in part due to their low mass.  Finally, there does not
seem to be any correlation between halo mass or mass fraction and rotation
speed alone.

However, at this point it is not unreasonable to conclude for NGC 5746 that
the three non-counterrotating clouds in Table 3 may just as likely be due to a
relatively inactive disk-halo cycle as to infall.  One possible further clue
to their origin would be infrared emission, and indeed imaging observations of
this and other edge-ons with the {\it Spitzer Space Telescope} have been
proposed.  A detection of dust would argue against an origin in low-metallicity
infalling gas.

A larger sample of edge-ons needs to be observed with comparable sensitivity
and angular resolution to understand how the mass and kinematics (rotational
lags in particular) of neutral halo gas relates to other galactic properties
such as star formation activity, assuming that such gas can always be
kinematically distinguished from warps and flares.  A better measure of such
activity than the somewhat crude tracer employed here would be the 24$\mu$m
surface density as measured by the {\it Spitzer Space Telescope}
\citep{2005ApJ...633..871C}, hence 24$\mu$m maps of a large sample of edge-ons
would be of great benefit in such comparisons.  For NGC 5746, deeper
observations with higher spectral resolution are clearly called for in order
to carry out a more complete census of halo gas that can be distinguished from
the warp, so that its spatial distribution, kinematics, and cloud masses can
be compared to the various models discussed here.

It should also be remembered that NGC 5746 is in a group environment, and some
of the high-latitude gas may owe its origin to previous encounters, although
the apparently low metallicity of the hot gas argues against encounters with
other large galaxies.

We reemphasize that our best warp model (Model B) is not necessarily
a unique fit to the data, especially because of asymmetries, but does
demonstrate that a warp is a superior explanation to a lagging halo.
It is not surprising that a warp is present in NGC 5746, as they are
common in disk galaxies \citep[e.g][]{2002A&A...394..769G}.  As far as
the warp parameters are concerned, we note that it is generally the case in
such galaxies \citep{1990ApJ...352...15B} that the line of nodes increases
in the direction of galaxy rotation for radii beyond the Holmberg radius,
which is 4.5' for NGC 5746 \citep{1958MeLu2.136....1H}.  Our warp model
implies an increase beginning just inside this radius.  More recently,
\citet{2007A&A...466..883V} finds evidence from edge-ons that warps begin just
beyond the radius where a truncation is evident in the stellar disk, but finds
no indication of such a truncation in NGC 5746.

Finally, there is the possibility that the warp itself is caused by infall.
The origin of warps is debated, but the possibility receiving most attention
recently is that they are due to infalling gas with an angular momentum vector
tilted with respect to the inner disk
\citep{1999MNRAS.303L...7J,2006MNRAS.370....2S}.  In the N-body simulations of
\citet{2006MNRAS.370....2S}, in which particles are injected into an outer
torus inclined at a fixed angle with respect to the main disk, the warp can
persist for several Gyr, making it difficult to infer the accretion history
from the mere presence of a warp.  That is, the large gas mass in the warp may
reflect accretion over a substantial period of time and not necessarily demand
a high accretion rate, as would more likely be the case such a large mass were
contained in a lagging halo.  On the other hand, if the centrally concentrated
X-ray halo emission reflects the hot gas density, then one would expect any
warm clouds recently formed by cooling out of the hot phase to be more
centrally concentrated.

\section{Conclusions}

We have observed the massive edge-on spiral NGC 5746 with the VLA, in an
attempt to find a vertically extended component of atomic gas as predicted by
recent galaxy formation models.  A high-latitude component has been
discovered, but almost all of its mass, summing to $1.2-1.6 \times 10^9$
\msun\ -- or about 15\% of the total HI mass -- is more readily explained as a
warp than as a halo.  The warp must be asymmetric in its column density
distribution and geometry, but this is not unusual.  If the high-latitude
component is a lagging halo, it must feature a large mass in a rather narrow
radial range, be centered at a large radius, and have a much larger degree of
asymmetry than the warp.  Even then, such a model has difficulty reproducing
the narrowness of the high-latitude spectra as well as the warp model.  We
therefore conclude that a warp is the more likely explanation.  The warp
itself may be a result of infall according to recent models.  It partially
resolves into clouds of mass $4 \times 10^6$ \msun\ to $10^8$ \msun,
accounting for about half the mass in the warp.

We have found four high-latitude features at velocities distinct from the
warp, totaling about 10$^8$ \msun.  These could be accreting onto the disk, or
they may originate in a disk-halo flow, although there is no other sign of
such a flow in this galaxy with little star forming activity.  One cloud is
counter-rotating and must have an external origin.  The cloud properties are
roughly comparable to those expected in the galaxy formation models of
\citet{2006MNRAS.370.1612K} and \citet{2006ApJ...644L...1S} where warm clouds
form by thermal instabilities in the residual hot halos.  However, these
models have not been tailored to galaxies as massive as NGC 5746, and it would
be interesting to compare our results with such a model.  Given the calculated
pressure in the hot halo, we expect any clouds to be primarily neutral.
Alternatively, we cannot rule out a weak disk-halo flow as the origin for
these clouds, except for the counter-rotating one.  How neutral gaseous halos
relate to other galaxian properties needs to be tested with more observations.

The disk shows the signature of radial inflow at the level of about
10 km s$^{-1}$.  This is most likely due to the bar, which is known to
have a strong kinematical signature from optical long-slit spectra.

The group member NGC 5740 is also detected, albeit outside the half-power
points of the primary beam.  The most important result here is a broad
extension of gas to the NE, which is more likely to be due to ram pressure
than tidal stripping, although whether the condition for such stripping is
met is unknown.


\acknowledgments

We thank J. Mulchaey for discussions regarding X-rays in groups, and L.
Sparke for discussion about warps.  We also thank an anonymous referee for
many useful comments.


The VLA is operated by the National Radio Astronomy Observatory, which is a
facility of the National Science Foundation operated under cooperative
agreement by Associated Universities, Inc.  We thank the VLA staff for making
these observations possible.  This research has made use of the NASA/IPAC
Extragalactic Database (NED) which is operated by the Jet Propulsion
Laboratory, California Institute of Technology, under contract with the
National Aeronautics and Space Administration.



Facilities: \facility{VLA}.




\bibliography{ms}

\clearpage





\begin{deluxetable}{cccccc}
\tabletypesize{\scriptsize}
\tablecaption{Models of the High Latitude Emission\label{tbl-1}}
\tablewidth{0pt}
\tablehead{
\colhead{Model} &
\colhead{Description} &
\colhead{Incl.}   &
\colhead{P.A.} &
\colhead{Disk rotation}   &
\colhead{High-latitude} \cr
\colhead{} &
\colhead{} &
\colhead{(degrees)} &
\colhead{(degrees)} &
\colhead{curve} &
\colhead{N$_{\rm H}$ profile}
}
\startdata
A & warp along    & Figure 11 & 350 & Figure 8 & Figure 6 \\
  & line of sight &           &     &          &          \\
B & warp at angle to & Figure 11 & Figure 11 & Figure 8 & Figure 6 \\
  & line of sight &           &     &          &          \\
C & lagging halo & 86.25 & 350 & Figure 8 & Figure 6 \\
  & (radial profile of disk) &           &     &          &          \\
D & lagging halo & 86.25 & 350 & Figure 8 & Figure 6 \\
  & (narrow ring) &           &     &          &          \\
E & lagging halo & 86.25 & 350 & Figure 8 & Figure 6 \\
  & (broader ring) &           &     &          &          \\
\enddata

 
\end{deluxetable}

\begin{deluxetable}{cccccc}
\tabletypesize{\scriptsize}
\tablecaption{Properties of Clumps in Extended Component\label{tbl-2}}
\tablewidth{0pt}
\tablehead{
\colhead{Number\tablenotemark{a}} &
\colhead{R.A.} &
\colhead{Decl.}   &
\colhead{Offset along major axis\tablenotemark{b}} &
\colhead{Offset along minor axis\tablenotemark{b}}   &
\colhead{Mass } \cr
\colhead{} &
\colhead{(J2000.0)} &
\colhead{(J2000.0)} &
\colhead{(arcmin)} &
\colhead{(arcmin)} &
\colhead{(10$^6$ \msun)}
}
\startdata
1  & 14$^{\rm h}$ 44$^{\rm m}$ 53.5$^{\rm s}$ & 2$^\circ$ 2' 13'' &  4.95& -0.25  &39  \\
2  & 14$^{\rm h}$ 44$^{\rm m}$ 56.0$^{\rm s}$ & 2$^\circ$ 1' 48''  &  4.45& -0.75  &7.4 \\
3  & 14$^{\rm h}$ 44$^{\rm m}$ 57.0$^{\rm s}$ & 2$^\circ$ 1' 21''  &  4.00& -0.85  &14  \\
4  & 14$^{\rm h}$ 44$^{\rm m}$ 56.9$^{\rm s}$ & 2$^\circ$ 0' 56''  &  3.60& -0.85  &7.9 \\
5  & 14$^{\rm h}$ 44$^{\rm m}$ 57.1$^{\rm s}$ & 2$^\circ$ 0' 12''  &  2.90& -0.75  &14  \\
6  & 14$^{\rm h}$ 44$^{\rm m}$ 57.3$^{\rm s}$ & 1$^\circ$ 59' 44''  &  2.45& -0.75  &15  \\
7  & 14$^{\rm h}$ 44$^{\rm m}$ 59.0$^{\rm s}$ & 1$^\circ$ 59' 14'' &  1.90& -1.10  &7.6 \\
8  & 14$^{\rm h}$ 44$^{\rm m}$ 57.8$^{\rm s}$ & 1$^\circ$ 58' 32'' &  1.15& -0.75  &31  \\
9  & 14$^{\rm h}$ 45$^{\rm m}$ 0.2$^{\rm s}$  & 1$^\circ$ 57' 24'' & -0.05& -1.05  &6.8 \\
10 & 14$^{\rm h}$ 44$^{\rm m}$ 59.6$^{\rm s}$ & 1$^\circ$ 56' 36''& -0.80& -0.75  &16  \\
11 & 14$^{\rm h}$ 45$^{\rm m}$ 1.0$^{\rm s}$ &  1$^\circ$ 55' 8''& -2.30& -0.85  &36  \\
12 & 14$^{\rm h}$ 45$^{\rm m}$ 1.4$^{\rm s}$ &  1$^\circ$ 53' 52''& -3.55& -0.70  &35  \\
13 & 14$^{\rm h}$ 45$^{\rm m}$ 2.7$^{\rm s}$ &  1$^\circ$ 53' 30''& -4.00& -1.00  &12 \\
14 & 14$^{\rm h}$ 45$^{\rm m}$ 2.9$^{\rm s}$ &  1$^\circ$ 52' 0''& -5.55& -0.75  &14  \\
15 & 14$^{\rm h}$ 44$^{\rm m}$ 59.2$^{\rm s}$ & 1$^\circ$ 52' 1''& -5.25&  0.15  &14  \\
16 & 14$^{\rm h}$ 44$^{\rm m}$ 58.4$^{\rm s}$ & 1$^\circ$ 53' 2''& -4.25&  0.10  &40  \\
17 & 14$^{\rm h}$ 44$^{\rm m}$ 57.2$^{\rm s}$ & 1$^\circ$ 53' 25''& -3.85&  0.35  &43  \\
18 & 14$^{\rm h}$ 44$^{\rm m}$ 55.9$^{\rm s}$  &1$^\circ$ 54' 8'' & -3.15&  0.45  &84  \\
19 & 14$^{\rm h}$ 44$^{\rm m}$ 53.8$^{\rm s}$  &1$^\circ$ 54' 1'' & -2.80&  1.00  &3.7 \\
20 & 14$^{\rm h}$ 44$^{\rm m}$ 54.0$^{\rm s}$  &1$^\circ$ 56' 28'' & -0.70&  0.60  &31  \\
21 & 14$^{\rm h}$ 44$^{\rm m}$ 52.4$^{\rm s}$  &1$^\circ$ 57' 22''  &  0.20&  0.85  &23  \\
22 & 14$^{\rm h}$ 44$^{\rm m}$ 52.5$^{\rm s}$  &1$^\circ$ 58' 8''  &  1.10&  0.75  &39  \\
23 & 14$^{\rm h}$ 44$^{\rm m}$ 51.6$^{\rm s}$  &1$^\circ$ 59' 2''  &  1.95&  0.80  &31  \\
24 & 14$^{\rm h}$ 44$^{\rm m}$ 49.7$^{\rm s}$  &1$^\circ$ 59' 19''  &  2.30&  1.20  &14  \\
25 & 14$^{\rm h}$ 44$^{\rm m}$ 51.9$^{\rm s}$  &1$^\circ$ 59' 56''  &  2.75&  0.50  &41  \\
26 & 14$^{\rm h}$ 44$^{\rm m}$ 50.1$^{\rm s}$  &1$^\circ$ 59' 55''  &  2.90&  1.00  &12  \\
27 & 14$^{\rm h}$ 44$^{\rm m}$ 50.2$^{\rm s}$  &2$^\circ$ 0'  14'' &  3.20&  0.95  &9.5 \\
28 & 14$^{\rm h}$ 44$^{\rm m}$ 50.2$^{\rm s}$  &2$^\circ$ 0'  42'' &  4.55&  0.50  &27  \\
29 & 14$^{\rm h}$ 44$^{\rm m}$ 50.8$^{\rm s}$  &2$^\circ$ 1'  42'' &  4.85&  0.90  &12  \\
30 & 14$^{\rm h}$ 44$^{\rm m}$ 49.0$^{\rm s}$  &2$^\circ$ 1'  54'' &  4.85&  0.90  &13  \\
\enddata

\tablenotetext{a}{Numbers generally run in a counter-clockwise direction around the disk
starting at the northern end.}

\tablenotetext{b}{Offsets are from the dynamical center determined in the text.}

 
\end{deluxetable}

\begin{deluxetable}{ccccccc}
\tabletypesize{\scriptsize}
\tablecaption{High-{\it z} Emission Features Not Associated with the Warp\label{tbl-3}}
\tablewidth{0pt}
\tablehead{
\colhead{Number} &
\colhead{R.A.} &
\colhead{Decl.}   &
\colhead{Offset along major axis\tablenotemark{a}} &
\colhead{Offset along minor axis\tablenotemark{a}}   &
\colhead{Mass} &
\colhead{Range in $V_{hel}$} \cr
\colhead{} &
\colhead{(J2000.0)} &
\colhead{(J2000.0)} &
\colhead{(arcmin)} &
\colhead{(arcmin)} &
\colhead{(10$^6$ \msun)} &
\colhead{(km s$^{-1}$)}
}
\startdata
1 & 14$^{\rm h}$ 44$^{\rm m}$ 56.6$^{\rm s}$ & 1$^\circ$ 59' 19'' &   2.00 & -0.45 &    7.2 & 1933--2017 \\
2 & 14$^{\rm h}$ 44$^{\rm m}$ 52.2$^{\rm s}$ & 1$^\circ$ 58' 43'' &   1.70 &  0.75 &    24  & 1493--1808 \\
3 & 14$^{\rm h}$ 44$^{\rm m}$ 58.0$^{\rm s}$ & 1$^\circ$ 58' 34'' &   1.15 & -0.70 &    56  & 1933--2058 \\
4 & 14$^{\rm h}$ 44$^{\rm m}$ 56.0$^{\rm s}$ & 1$^\circ$ 54' 19'' &  -2.90 &  0.60 &    7.9 & 1391--1495 \\
\enddata

\tablenotetext{a}{Offsets are from the dynamical center determined in the text.}

 
\end{deluxetable}

\begin{deluxetable}{ccccc}
\tabletypesize{\scriptsize}
\tablecaption{Neutral Gaseous Halos of Nearby Galaxies\label{tbl-4}}
\tablewidth{0pt}
\tablehead{
\colhead{Galaxy} &
\colhead{HI halo mass} &
\colhead{Percentage of HI} &
\colhead{$L_{FIR}$/$D_{25}^2$}   &
\colhead{Rotation speed\tablenotemark{a}} \\
\colhead{} &
\colhead{($10^8$ \msun)} &
\colhead{mass in halo} &
\colhead{($10^{40}$ erg s$^{-1}$ kpc$^{-2}$)} &
\colhead{(km s$^{-1}$)}\\
}
\startdata
NGC 891\tablenotemark{b}   & 12  & 30	&   3.5 & 225 \\
NGC 4559\tablenotemark{c}  &  6  & 10	&   0.8 & 120 \\
NGC 3044\tablenotemark{d}  &  4  & 7    &   4.7 & 150 \\
NGC 2403\tablenotemark{e}  &  3  & 10	&   1.1 & 135 \\
NGC 6946\tablenotemark{f}  &  3  & 4    &   9.6 & 170 \\
UGC 7321\tablenotemark{g}  & 1.3 & 12	&   0.1 & 105 \\
NGC 5746                   & 1.0 & 1    &   0.4 & 310 \\
NGC 253\tablenotemark{h}   & 0.8 & 3	&  10.6 & 230 \\
M31\tablenotemark{i}   & 0.35 & 1	&   0.2 & 270 \\
\enddata

\tablenotetext{a}{An approximate value for the flat part of the disk rotation
curve.}
\tablenotetext{b}{Halo mass and rotation speed from \citet{2007AJ....134.1019O}} 
\tablenotetext{c}{Halo mass and rotation speed from \citet{2005A&A...439..947B}}
\tablenotetext{d}{Halo mass and rotation speed from \citet{1997ApJ...490..247L}} 
\tablenotetext{e}{Halo mass and rotation speed from \citet{2002AJ....123.3124F}} 
\tablenotetext{f}{Halo mass and rotation speed from \citet{2007PhDT.........1B}} 
\tablenotetext{g}{Halo mass from \citet{2003ApJ...593..721M} and rotation speed
from \citet{2003AJ....125.2455U}}
\tablenotetext{h}{Halo mass and rotation speed from \citet{2005A&A...431...65B}}
\tablenotetext{i}{Halo mass from \citet{2004ApJ...601L..39T}, total mass from 
\citet{1992ApJ...386..120B},
rotation speed from \citet{1984A&A...141..195B}}

 
\end{deluxetable}

\end{document}